\documentclass[fleqn,usenatbib]{mnras}

\usepackage{newtxtext,newtxmath}
\usepackage{mathrsfs}

\usepackage[T1]{fontenc}
\usepackage{ae,aecompl}
\usepackage{booktabs}

\usepackage{graphicx}	
\usepackage{amsmath}	
\usepackage{amssymb}	

\usepackage{empheq}

\title[No evolution of non-repeating FRB rates]{No redshift evolution of non-repeating fast radio-burst rates}

\author[T. Hashimoto et al.]{
Tetsuya Hashimoto,$^{1,2}$\thanks{E-mail: tetsuya@phys.nthu.edu.tw}
Tomotsugu Goto,$^{1}$
Alvina Y. L. On,$^{1,2,3}$
Ting-Yi Lu,$^{1}$
\newauthor
Daryl Joe D. Santos,$^{1}$
Simon C.-C. Ho,$^{1}$
Seong Jin Kim,$^{1}$
Ting-Wen Wang,$^{1}$
\newauthor
and Tiger Y.-Y. Hsiao$^{1}$
\\
$^{1}$Institute of Astronomy, National Tsing Hua University, 101, Section 2. Kuang-Fu Road, Hsinchu, 30013, Taiwan (R.O.C.)\\
$^{2}$Centre for Informatics and Computation in Astronomy (CICA), National Tsing Hua University,\\
101, Section 2. Kuang-Fu Road, Hsinchu, 30013, Taiwan (R.O.C.)\\
$^{3}$Mullard Space Science Laboratory, University College London, Holmbury St Mary, Surrey RH5 6NT, UK
}

\date{Accepted 2020 August 14. Received 2020 July 29; in original form 2020 May 19}

\pubyear{2020}

\begin{document}
\label{firstpage}
\pagerange{\pageref{firstpage}--\pageref{lastpage}}
\maketitle

\begin{abstract} 
Fast radio bursts (FRBs) are millisecond transients of unknown origin(s) occurring at cosmological distances.
Here we, for the first time, show time-integrated-luminosity functions and volumetric occurrence rates of non-repeating and repeating FRBs against redshift.
The time-integrated-luminosity functions of non-repeating FRBs do not show any significant redshift evolution.
The volumetric occurrence rates are almost constant during the past $\sim$10 Gyr.
The nearly-constant rate is consistent with a flat trend of cosmic stellar-mass density traced by old stellar populations.
Our findings indicate that the occurrence rate of non-repeating FRBs follows the stellar-mass evolution of long-living objects with $\sim$Gyr time scales, favouring e.g. white dwarfs, neutron stars, and black holes, as likely progenitors of non-repeating FRBs.
In contrast, the occurrence rates of repeating FRBs may increase towards higher redshifts in a similar way to the cosmic star formation-rate density or black hole accretion-rate density if the slope of their luminosity function does not evolve with redshift.
Short-living objects with $\lesssim$ Myr time scales associated with young stellar populations (or their remnants, e.g., supernova remnants, young pulsars, and magnetars) or active galactic nuclei might be favoured as progenitor candidates of repeating FRBs.
\end{abstract}




\section{Introduction}
\label{introduction}
Since the first discovery of fast radio burst (FRB) \citep{Lorimer2007}, more than 100 FRBs have been detected \citep{Petroff2016}.
They are usually classified into two populations of non-repeating and repeating FRBs.
The nature of an FRB being a repeater or non-repeater may not be intrinsic but instead observational. 
Therefore, there are likely to be some \lq non-repeaters\rq\ that are actually repeaters in the current FRB sample.
Numerous theoretical models of FRB progenitors have been proposed to date for non-repeating and repeating FRBs \citep[e.g., ][]{Platts2019}.
However, the origin(s) is still unknown.
A unique observable of FRB is a dispersion measure.
The dispersion measure is a time lag of burst arrival depending on observed frequency, which tells us how much ionised materials exist along the line of sight between the FRB and the observer \citep[e.g., ][]{Thornton2013}.
By modelling and removing the dispersion measure contributions from the Milky Way and host galaxies, the dispersion measure of intervening intergalactic medium (IGM) can be utilised as an indicator of redshift or distance to the FRB.
Therefore, a \lq time-integrated luminosity\rq\ can be calculated using an observed fluence and distance to the FRB \citep{Hashimoto2019}, where the time-integrated luminosity is the luminosity integrated over the duration of the FRB.
We define the time-integrated luminosity for individual bursts of each repeating-FRB source (and not to be confused with the integral over multiple bursts of repeaters).
The increasing statistics of FRBs allow us to estimate their time-integrated-luminosity functions (hereafter luminosity functions) and volumetric occurrence rates.
These are defined as the number densities of FRBs per unit time-integrated luminosity per unit time (i.e., luminosity-dependent volumetric rate) and the luminosity-integrated luminosity functions, respectively.

\citet{Ravi2019} calculated the burst occurrence rate for both non-repeating and repeating FRBs using only a few closest events, i.e., two or three FRBs with the lowest dispersion measures.
However, such occurrence rate would suffer from a large uncertainty.
The observed dispersion measures of such FRBs are relatively more contaminated by the Milky Way and host galaxy contributions, since the dispersion measure of IGM, i.e., the distance indicator, is much smaller than the contamination (see Section \ref{calcz} for details). 
Hence, the measurements of distance and number density of very nearby FRBs are more uncertain.
The volumetric occurrence rate calculated from very nearby FRBs is also sensitive to the peculiarities of the few lowest dispersion-measure events, and is limited to the measurement at the local Universe.
The number density of FRBs strongly depends on the luminosity or time-integrated luminosity \citep[e.g., ][]{Luo2018,Luo2020,Hashimoto2020}, which was not taken into account in \citet{Ravi2019}.
In this work, we exclude very nearby FRBs which have a large uncertainty and take the luminosity dependency of FRB number density into account.
The luminosity functions allow us to investigate the redshift evolution of volumetric occurrence rates of FRBs.

The structure of the paper is as follows:
we describe an FRB catalogue used in this work in Section \ref{catalog}.
In Section \ref{analysis}, we demonstrate calculations of redshift and luminosity functions as well as sample selection criteria for the luminosity functions.
Results of the luminosity functions and volumetric occurrence rates are described in Section \ref{results}.
The indications of our results on non-repeating and repeating FRB populations and their origins are discussed in Section \ref{discussion} followed by conclusions in Section \ref{conclusion}.

Throughout this paper, we assume the {\it Planck15} cosmology \citep{Planck2016} as a fiducial model, i.e., $\Lambda$ cold dark matter cosmology with ($\Omega_{m}$,$\Omega_{\Lambda}$,$\Omega_{b}$,$h$)=(0.307, 0.693, 0.0486, 0.677), unless otherwise mentioned.

\section{Catalogue description}
\label{catalog}
In this work we use the updated version of an FRB catalogue\footnote[1]{\url{http://www.phys.nthu.edu.tw/~tetsuya/}} constructed in a previous study \citep{Hashimoto2020}.
The catalogue is composed of information from the Fast Radio Burst Catalogue (FRBCAT) project \citep{Petroff2016} as of 24 Feb. 2020 along with complementary information on individual bursts of repeating FRBs \citep{Spitler2016,Scholz2016,Zhang2018,CHIMEFRB2019,CHIME1repeat2019,CHIME8repeat2019,Kumar2019,Fonseca2020}.
FRBs in the catalogue are \lq verified\rq\ through publications or high importance scores in the VOEvent Network \citep{Petroff2017}.
The catalogue includes instrumental parameters and observed quantities of FRBs, i.e., FRB ID, telescope, galactic longitude ($l$), latitude ($b$), sampling time ($w_{\rm sample}$), central frequency ($\nu_{\rm obs}$), observed dispersion measure (DM$_{\rm obs}$), observed burst duration ($w_{\rm obs}$), observed fluence ($E_{\nu_{\rm obs}}$), and spectral index ($\alpha$), as well as errors of these observed parameters \citep[see][for details]{Hashimoto2020}.
FRBs with dispersion measures dominated by the Milky-Way and host-galaxy contributions are excluded from the catalogue if the spectroscopic redshift is not available, i.e., DM$_{\rm obs}-$DM$_{\rm MW}-$DM$_{\rm halo}-$DM$_{\rm host}\leq$0, where DM$_{\rm MW}$, DM$_{\rm halo}$, and DM$_{\rm host}$ are dispersion measures contributed from the interstellar medium in the Milky Way, the dark matter halo hosting the Milky Way, and the galaxy hosting FRB, respectively. 
This is because the uncertainties of redshifts and distances of such FRBs are too large to calculate the luminosities as mentioned in Main section.
In summary, the catalogue contains a total of 84 non-repeating FRBs and 19 repeating FRBs with 164 repeats.

The properties of FRBs selected in Section \ref{selection} are summarised in Tables \ref{tabA1} and \ref{tabA2} in APPENDIX A.
For every individual repeating burst, its physical parameters are calculated, satisfying the selection criteria in Section \ref{selection}. 
The median over the individual bursts is then assigned to parameterise each source of repeating FRBs (see Tables \ref{tabA1} and \ref{tabA2}).

\section{Analysis and sample selection}
\label{analysis}
\subsection{Calculation of redshift}
\label{calcz}
Spectroscopic redshifts are available for four FRB host galaxies in the catalogue as of 24 Feb. 2020.
For other FRBs, redshifts are calculated from DM$_{\rm obs}$ which is composed of DM$_{\rm MW}$, DM$_{\rm halo}$, DM$_{\rm IGM}$ and DM$_{\rm host}$, i.e.,
\begin{equation}
\label{eqDM}
{\rm DM}_{\rm obs}= {\rm DM}_{\rm MW}(b,l)+{\rm DM}_{\rm halo}+{\rm DM}_{\rm IGM}(z)+{\rm DM}_{\rm host}(z).
\end{equation}
Here DM$_{\rm IGM}$ is the dispersion measure contributed from the intervening IGM \citep[see][for details]{Hashimoto2019}.
We use the YMW16 \citep{Yao2017} electron-density model of the Milky Way to estimate DM$_{\rm MW}$.
We integrate DM$_{\rm MW}$ up to a distance of 10 kpc along the line of sight to the FRB. 
We adopt DM$_{\rm halo}=65$ pc cm$^{-1}$ which is an average value reported in literature \citep{Prochaska2019DMhalo}.
The DM$_{\rm halo}$ is estimated to be between 10 kpc and 200 kpc from the Sun.
Following previous study \citep{Shannon2018}, DM$_{\rm host}$ is parameterised as 
\begin{equation}
\label{eqDMhost}
{\rm DM}_{\rm host}=50.0/(1+z) {\rm \ pc\ cm}^{-3}.
\end{equation}

On average, intervening ionised materials increase with increasing distance to the FRB.
Therefore, DM$_{\rm IGM}$ is an indicator of distance and redshift to the FRB.
The averaged DM$_{\rm IGM}$ is analytically expressed as a function of redshift with assumed cosmological parameters and IGM evolution \citep{Zhou2014}, i.e.,
\begin{equation}
\label{eqDMIGM}
\begin{split}
&\mathrm{DM}_{\mathrm{IGM}}(z)=\Omega_{\mathrm{b}} \frac{3 H_{0} c}{8 \pi G m_{\mathrm{p}}} \times \\
&\int_{0}^{z} \frac{\left(1+z^{\prime}\right) f_{\mathrm{IGM}}\left(z^{\prime}\right)\left(Y_{\mathrm{H}} X_{\mathrm{e}, \mathrm{H}}\left(z^{\prime}\right)+\frac{1}{2} Y_{\mathrm{p}} X_{\mathrm{e}, \mathrm{He}}\left(z^{\prime}\right)\right)}{\left\{\Omega_{\mathrm{m}}\left(1+z^{\prime}\right)^{3}+\Omega_{\Lambda}\left(1+z^{\prime}\right)^{3\left[1+w\left(z^{\prime}\right)\right]}\right\}^{1 / 2}} d z^{\prime}
\end{split}
\end{equation}
for a flat Universe.
Here $X_{\mathrm{e}, \mathrm{H}}$ and $X_{\mathrm{e}, \mathrm{He}}$ are the ionisation fractions of the intergalactic hydrogen and helium, respectively.
$Y_{\mathrm{H}}=\frac{3}{4}$ and $Y_{\mathrm{p}}=\frac{1}{4}$ are the mass fractions of H and He.
$f_{\mathrm{IGM}}$ is the fraction of baryons in the IGM.
The equation of state of dark energy is expressed as $w$, where $w=-1$ is assumed for the constant dark energy \citep{Chevallier2001,Linder2003}.
We assumed $X_{\mathrm{e}, \mathrm{H}}=1$ and $X_{\mathrm{e}, \mathrm{He}}=1$, which are reasonable up to $z \sim 3$ because the IGM is almost fully ionised.
We adopted $f_{\mathrm{IGM}}$ = 0.9 at $z>1.5$ and $f_{\mathrm{IGM}}= 0.053z+0.82$ at $z\leq1.5$ following literature \citep{Zhou2014}.
Taking equations (\ref{eqDMhost}) and (\ref{eqDMIGM}) into account, the right term of equation (\ref{eqDM}) is a function of redshift.
The solution provides an individual FRB with a redshift measurement.
Since each repeating FRB with the same ID has multiple bursts, the redshifts are individually calculated for multiple bursts of each repeater.

The redshift uncertainty of each FRB is calculated by independently assigning 10,000 random errors on DM$_{\rm obs}$ and DM$_{\rm IGM}$.
The random errors on DM$_{\rm obs}$ are assumed to follow Gaussian probability distribution functions with standard deviations of observational errors of DM$_{\rm obs}$.
The DM$_{\rm IGM}$ error is attributed to the inhomogeneity of the IGM along different lines of sight in the Universe. 
We use results from recent cosmological simulations \citep{Zhu2018}.
Among the simulations, a case with the largest fluctuation of DM$_{\rm IGM}$ is adopted as a conservative assumption (see APPENDIX B for details).
Similarly, the random errors following the IGM fluctuation are added to DM$_{\rm IGM}$ to calculate the redshift uncertainty.

\begin{figure}
    \includegraphics[width=\columnwidth]{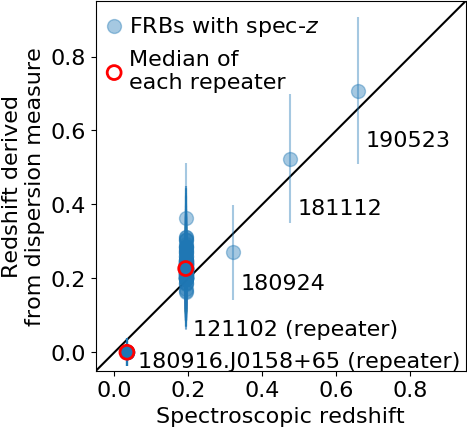}
    \caption{
    Comparison between spectroscopic and dispersion measure-derived redshifts in the catalogue.
    Repeating FRBs have multiple redshift measurements corresponding to individual repeated bursts.
    Red circles are median redshifts among repeated bursts of FRB121102 and 180916.J0158+65.
    }
    \label{fig1}
\end{figure}

In Fig. \ref{fig1}, spectroscopic and dispersion measure-derived redshifts are compared.
Since FRB 121102 and 180916.J0158+65 are repeating FRBs, the redshifts are calculated for individual repeated bursts.
Red circle indicates the median value of dispersion measure-derived redshifts for each repeater. 
We confirmed that spectroscopic and dispersion measure-derived redshifts are consistent within the uncertainties.
Recently, there has been an increasing number of FRB host galaxy identifications \citep{Macquart2020}.
However, these new localisations have not been included in FRBCAT, as of 24 Feb. 2020, which is used in this work. 
Therefore, we do not include them in our sample, but we note their importance in testing the accuracy of dispersion measure-derived redshifts.

The redshift distributions of non-repeating and repeating FRBs detected with each telescope in the catalogue are shown in Fig. \ref{fig2}.
In this work, we only consider non-repeating FRBs detected with Parkes and repeating FRBs detected with the Canadian Hydrogen Intensity Mapping Experiment (CHIME).
This is because these dominate the sample, covering wide ranges of redshifts compared to FRBs detected with other telescopes.

\subsection{Calculation of time-integrated luminosity}
\label{calcLnu}
Using the redshift of each FRB calculated in the previous section, time-integrated luminosity is estimated from the observed fluence with assumptions of cosmological parameters.
The time-integrated luminosity, $L_{\nu_{\rm rest}}$, is expressed as \citep{Novak2017}:
\begin{equation}
\label{eqLnu}
L_{\nu_{\rm rest}}=\frac{4\pi d_{l}(z)^{2}}{(1+z)^{2+\alpha}} \left(\frac{\nu_{\rm rest}}{\nu_{\rm obs}} \right)^{\alpha} E_{\nu_{\rm obs}},
\end{equation}
where d$_{l}(z)$ is a luminosity distance to the FRB calculated from redshift, and a power-law FRB spectrum expressed as $E_{\nu_{\rm obs}} \propto \nu_{\rm obs}^{\alpha}$ is assumed.
A typical observed spectral slope of $\alpha=-1.5$ \citep{Macquart2019} is adopted for FRBs without $\alpha$ measurement.
We use $\nu_{\rm rest}=1.83$ GHz, which minimises the $K$-correction term, $\frac{1}{(1+z)^{2+\alpha}}\left( \frac{\nu_{\rm rest}}{\nu_{\rm obs}} \right)^{\alpha}$.
Since each repeating FRB with the same ID has multiple bursts, the time-integrated luminosities are individually calculated for multiple bursts of each repeater.

\subsection{Detection threshold}
In previous studies, detection thresholds on observed fluence, $E_{\rm lim}$, were reported for Parkes and CHIME FRBs.
However, the definitions are slightly different. 
The reported CHIME detection threshold explicitly includes a duration dependency ($E_{\rm lim}\sim1.0w_{\rm obs}^{1/2}$) \citep{CHIMEFRB2019}, whereas Parkes does not include the duration dependency in \citet{Keane2015} but includes it in \citet{Keane2018}.
The different definitions of the detection threshold could cause additional systematics in the luminosity functions.
Following \citet{Hashimoto2020}, we empirically standardise the detection thresholds of Parkes and CHIME.
Panels (a) and (b) in Fig. \ref{fig3} demonstrate the observed duration as a function of observed fluence of FRBs detected with each telescope.
Note that all of the FRBs detected with Parkes and CHIME as of 24 Feb. 2020 are shown before applying any of criteria.
The dashed and dotted lines indicate detection thresholds reported in previous studies \citep{Keane2015,Keane2018,CHIMEFRB2019}.
In this work, the values of $E_{\rm lim}$ for Parkes and CHIME are approximated by the peaks of the data distributions along the perpendicular direction to the $w_{\rm obs}^{1/2}$ dependency in panels (a) and (b) in Fig. \ref{fig3}, such that the duration dependency can be taken into account.
The peak is adopted to reduce the uncertainty of detection completeness.
To investigate the peak of data distribution, panels (a) and (b) in Fig. \ref{fig3} were rotated so that the $w_{\rm obs}^{1/2}$ dependency can be aligned along the vertical axis (panels c and d in Fig. \ref{fig3}).
The peaks of histograms in Fig. \ref{fig3} correspond to empirically determined $E_{\rm lim}$ as shown by the black solid lines.
The derived thresholds are $E_{\rm lim}=0.72w_{\rm obs}^{1/2}$ and $1.4w_{\rm obs}^{1/2}$ (Jy ms) for Parkes and CHIME, respectively.
We confirmed that the empirically determined $E_{\rm lim}$ for Parkes and CHIME (black solid lines in Fig. \ref{fig3}) are almost the same as the thresholds reported by \citet{Keane2018} (dotted lines in panels a and c: $E_{\nu}=0.50w_{\rm obs}^{1/2}$ Jy ms) and \citet{CHIMEFRB2019} (dashed lines in panels b and d: $E_{\nu}=1.0w_{\rm obs}^{1/2}$ Jy ms), respectively. 

The detection threshold is applied to all observed fluences and durations regardless of the source redshift. 
A broader pulse is relatively difficult to detect due to (i) broad intrinsic duration, (ii) scattering broadening between the observer and FRB progenitor, (iii) redshift time dilation, and/or (iv) instrumental broadening. 
Since we consider the duration-dependent detection threshold using observed durations of individual sources, these effects are taken into account when computing survey volumes (see the following sections).

\subsection{Sample selection for luminosity functions}
\label{selection}
\begin{figure*}
    \includegraphics[width=\columnwidth]{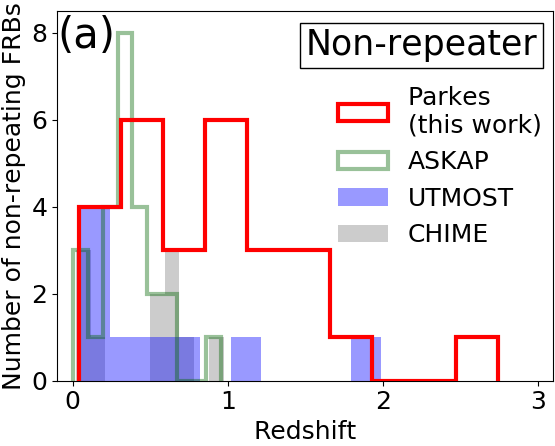}
    \includegraphics[width=\columnwidth]{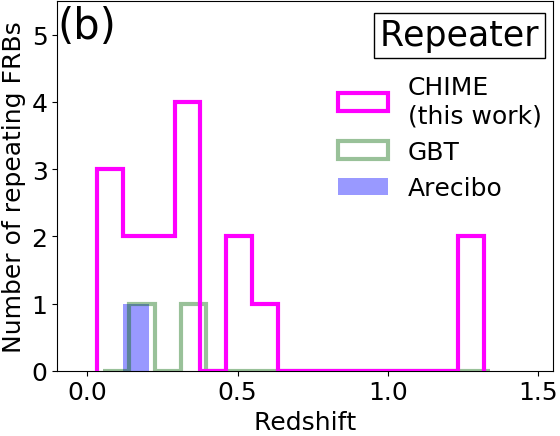}
    \includegraphics[width=\columnwidth]{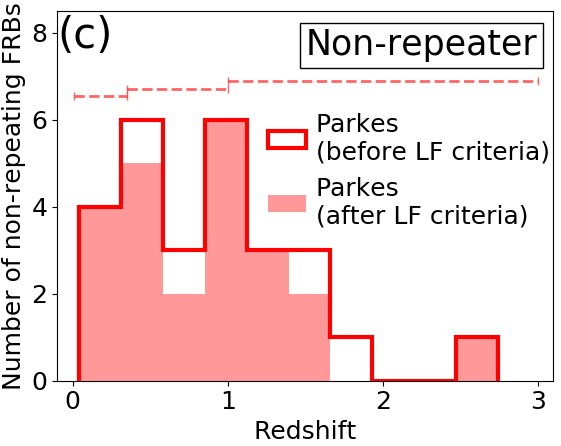}
    \includegraphics[width=\columnwidth]{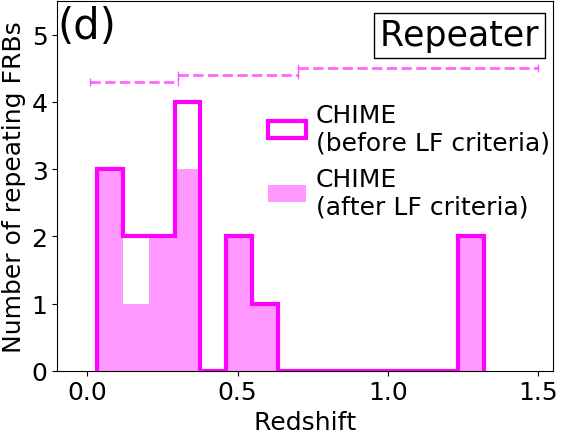}
    \caption{
    Redshift distributions of FRBs.
    Spectroscopic redshift is used for the histograms if available.
    Otherwise, redshift is derived from the dispersion measure.
    (a) Non-repeating and (b) repeating FRBs detected by each telescope in the catalogue. 
    Different colours correspond to different telescopes including Parkes, ASKAP, UTMOST, and CHIME for non-repeating FRBs and CHIME, GBT, and Arecibo for repeating FRBs.
    The repeating FRBs are counted such that the identical FRB ID indicates the same source.
    For each repeating FRB ID, the redshift is calculated by a median value of its individual repeated bursts.
    (c) Redshift distributions of Parkes non-repeating FRBs before and after applying the criteria for luminosity function (LF), i.e., $z>0.01$ and $E_{\nu_{\rm obs}}\geq E_{\rm lim}$.
    The red shaded sample is utilised for the calculations of luminosity functions and volumetric occurrence rates.
    Horizontal dashed lines indicate redshift bins to calculate luminosity functions and volumetric occurrence rates.
    (d) Same as (c) except for CHIME repeating FRBs.
    }
    \label{fig2}
\end{figure*}

\begin{figure*}
    \includegraphics[width=2.0\columnwidth]{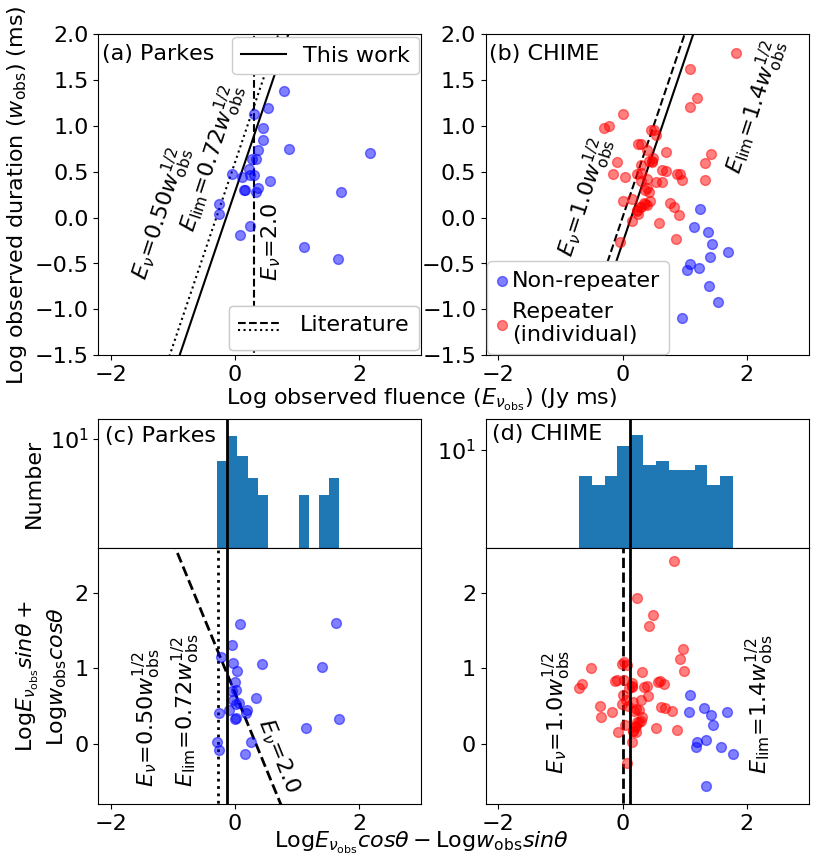}
    \caption{
    Empirically derived detection thresholds of Parkes and CHIME.
    Note that all of the FRBs detected with Parkes and CHIME as of 24 Feb. 2020 are shown before applying any of the criteria.
    In each panel, the solid line indicates the empirically derived detection threshold, $E_{\rm lim}$, in this work.
    The dashed and dotted lines correspond to the detection thresholds reported in the literature: dashed line from \citet{Keane2015} and dotted line from \citet{Keane2018} for Parkes FRBs, and dashed line from \citet{CHIMEFRB2019} for CHIME FRBs.
    (a) Observed duration of Parkes FRBs as a function of observed fluence.
    (b) Same as (a) except for FRBs detected with CHIME.
    Non-repeating and repeating FRBs are shown in different colours.
    Individual repeated bursts of repeaters are demonstrated.
    (c) Rotated version of (a) so that the duration dependency on $E_{\rm lim}$ can be aligned along the vertical axis together with a histogram.
    $E_{\rm lim}$ is approximated by the peak of the histogram.
    The rotation angle, $\theta$, is $\pi/2-\tan^{-1}(2.0)$ rad.
    (d) Same as (c) except for CHIME FRBs.
    }
    \label{fig3}
\end{figure*}

From the catalogue, we selected FRBs satisfying the following criteria to calculate the luminosity functions:
\begin{itemize}
\item Parkes (CHIME) detection for non-repeating (repeating) FRBs
\item Redshift (spec-$z$ if available) $>$ 0.01 
\item $E_{\nu_{\rm obs}}\geq E_{\rm lim}$
\end{itemize}
We only consider non-repeating FRBs detected with Parkes and repeating FRBs detected with CHIME, because these dominate the sample at the moment, covering wider ranges of redshifts compared to FRBs detected with other telescopes (Fig. \ref{fig2}).
For instance, Parkes sample includes FRBs at higher redshifts than those detected by ASKAP, allowing us to investigate the redshift evolution of luminosity functions beyond $z\sim1$.
The lower limit on redshift is applied to exclude very close events that could cause large uncertainties on the dispersion measure-derived distances.
The sample for luminosity functions consists of a total of 23 Parkes non-repeating FRBs and 14 CHIME repeating FRBs with 30 repeats.
The redshift distributions of samples before and after applying the redshift and $E_{\rm lim}$ criteria are shown in panels (c) and (d) in Fig. \ref{fig2}.
The redshifts and time-integrated luminosities are calculated for individual bursts of each source of repeating FRBs.
We adopt median values of the redshifts and time-integrated luminosities for each repeater after applying the above criteria.

\subsection{Calculation of luminosity functions}
\label{calcLF}
Following previous literature \citep[e.g.,][]{Schmidt1968,Avni1980,Hashimoto2020}, here we calculate FRB luminosity functions.
In our sample, non-repeating and repeating FRBs were divided into three different redshift bins.
The bins are $0.01<z\leq0.35$, $0.35<z\leq1.0$, and $1.0<z\leq3.0$ for non-repeating FRBs; and $0.01<z\leq0.3$, $0.3<z\leq0.7$, and $0.7<z\leq1.5$ for repeating FRBs.
The median redshifts of sample in these bins are $z_{\rm median}=0.16$, 0.76, and 1.3 for non-repeating FRBs and $z_{\rm median}=0.17$, 0.48, and 1.3 for repeating FRBs, respectively.
These bins are selected such that their time intervals are equal, corresponding to approximately 4 and 3 Gyr for non-repeating and repeating FRBs, respectively. 
The luminosity function of each redshift bin is independently calculated by a $V_{\rm max}$ method.
The 4$\pi$ coverage of $V_{\rm max}$, $V_{\rm max,4\pi}$, is calculated for individual FRBs according to 
\begin{equation}
\label{vmax}
V_{\rm max,4\pi}=\frac{4\pi}{3}(d_{\rm max}^{3}-d_{\rm min}^{3}),
\end{equation}
where $d_{\rm min}$ is the comoving distance at the lower bound of redshift bin to which the FRB belongs.
$d_{\rm max}$ is the maximum comoving distance for the FRB with $L_{\nu_{\rm rest}}$ to be detected with the detection threshold, $E_{\rm lim}$.
Using a conversion of $d_{l}(z)$ in equation (\ref{eqLnu}) to the comoving distance, $d_{\rm max}$ is expressed as
\begin{equation}
\label{eqdmax}
d_{\rm max} (z_{\rm max})=\left[\frac{L_{\nu_{\rm rest}}(1+z_{\rm max})^{2+\alpha}}{4\pi E_{\rm lim}}\left(\frac{\nu_{\rm obs}}{\nu_{\rm rest}}\right)^{\alpha}\right]^{1/2}(1+z_{\rm max})^{-1},
\end{equation}
where $z_{\rm max}$ is redshift at the comoving distance of $d_{\rm max}$.
Since the left term of equation (\ref{eqdmax}), comoving distance, is calculated as a function of redshift with a cosmological assumption, the solution to $z_{\rm max}$ of equation (\ref{eqdmax}) provides an individual FRB with $d_{\rm max}$.
If $z_{\rm max}$ is larger than the upper bound of the redshift bin to which the FRB belongs, the upper bound is utilised as $z_{\rm max}$ so that $z_{\rm max}$ cannot exceed the redshift limit.

Each FRB was detected in a comoving volume of $V_{\rm max,4\pi}\times \Omega_{\rm sky}$ during rest-frame survey time, $t_{\rm rest}=t_{\rm obs}/(1+z_{\rm FRB})$, where $\Omega_{\rm sky}$ and $z_{\rm FRB}$ are the fractional sky coverage of the survey and redshift of the FRB, respectively.
Therefore the number density of each FRB per unit time, $\rho (L_{\nu_{\rm rest}})$, is
\begin{equation}
\label{eqrho}
\rho(L_{\nu_{\rm rest}})=1/(V_{\rm max,4\pi}\Omega_{\rm sky}t_{\rm rest})=(1+z_{\rm FRB})/(V_{\rm max,4\pi}\Omega_{\rm sky}t_{\rm obs}).
\end{equation}
Each survey provides a different value of $\Omega_{\rm sky}t_{\rm obs}$.
When FRBs are detected via multiple surveys, $\Omega_{\rm sky}t_{\rm obs}$ was accumulated for each telescope as 
\begin{equation}
\label{eqeqrhosum}
\rho(L_{\nu_{\rm rest}})=(1+z_{\rm FRB})/(V_{\rm max,4\pi}\Sigma_{i}\Omega_{\rm sky, {\it i}}t_{\rm obs, {\it i}}),
\end{equation}
where $i$ denotes $i$th survey with the same telescope. The 23 Parkes non-repeating FRBs were detected via observations with survey areas of 267 and 4394.5 (deg$^{2}$ hour) \citep{Zhang2019,Oslowski2019}.
The 14 CHIME repeating FRBs were detected via observations with a survey area of 250 (deg$^{2}$) $\times$ 399 (day) $\times$ 24 (hour) \citep{Fonseca2020}.
Therefore, the adopted $\Sigma_{i}\Omega_{\rm sky, {\it i}}t_{\rm obs, {\it i}}$ are 4661.5 and 2.39$\times10^{6}$ (deg$^{2}$ hour) for Parkes and CHIME samples, respectively.

The non-repeating FRB sample at each redshift bin was divided into two luminosity bins, $L_{j} (j=1,2)$, in logarithmic scale. 
Here we select the two bins such that each bin includes at least two objects.
The repeating FRB sample at $0.7<z\leq1.5$ has only one luminosity bin due to the small sample size.
Within the luminosity bins, $\rho (L_{j})$ is summed to derive the luminosity function, $\Phi$, i.e., 
\begin{equation}
\label{LF}
\Phi (z_{\rm median}, L_{j})=\Sigma_{k}\rho(L_{j,k})/\Delta\log L,
\end{equation}
where the subscript $k$ denotes the $k$th FRB in $L_{j}$ bin and $\Delta\log L$ is the luminosity bin size.

Following the approach of e.g., \citet{Gabasch2004} and \citet{Caputi2007}, we added the Poisson error and errors of $\rho$ ($\delta\rho$) in quadrature in each luminosity bin to derive the uncertainty of $\Phi$ (vertical error bars in Fig. \ref{fig4}). 
For individual FRBs, $\delta\rho$ is calculated by Monte Carlo simulations with 10,000 iterations. 
In the simulations, observational errors of fluence, duration, dispersion measure, and the fluctuation of DM$_{\rm IGM}$ along different lines of sight propagate to Eq. \ref{eqeqrhosum}, thus providing individual FRBs with $\delta\rho$.
Fitting uncertainties to the luminosity functions are also estimated by Monte Carlo simulations.
In each simulation, random errors on $\log \Phi$ are independently assigned to the data points in Fig. \ref{fig4}.
The errors follow Gaussian probability distributions with standard deviations of the uncertainties of $\Phi$ derived above.
The randomised data points are fitted with a linear function to derive a slope and vertical intercept.
This process was iterated 10,000 times to derive $\pm$ 1 $\sigma$ uncertainties of the fit (dotted lines in Fig. \ref{fig4}).
Uncertainties of the volumetric occurrence rates in Fig. \ref{fig5} include these fitting uncertainties.

\section{Results}
\label{results}
\subsection{Luminosity functions}
\subsubsection{Non-repeating FRBs}
\label{z_evol_function}
In this work, to mitigate systematics in different telescopes, we only consider 23 non-repeating FRBs detected with the Parkes radio telescope and 14 repeating FRBs detected with the CHIME (see Section \ref{analysis} for details).
The luminosity functions of non-repeating FRBs at three different redshift bins are shown in Fig. \ref{fig4}a.
The linear best-fit functions are shown by the coloured thick-dashed lines, with 1$\sigma$ fitting uncertainties by the dotted lines.
These functions are 
\begin{equation*}
\log\Phi_{\rm NR}=
\end{equation*}
\vspace{-20pt}
\begin{empheq}[left={\empheqlbrace}]{alignat=2}
\label{LFnorepeat1}
& (-0.35^{+0.51}_{-0.50}) (\log L_{\nu} -31.0)+(4.3^{+0.2}_{-0.4})(0.01<z\leq0.35)\\
\label{LFnorepeat2}
& (-0.38^{+0.37}_{-0.50}) (\log L_{\nu} -31.0)+(4.3^{+0.3}_{-0.2})~(0.35<z\leq1.0)\\
\label{LFnorepeat3}
& (-0.43^{+1.19}_{-1.05}) (\log L_{\nu} -31.0)+(4.3^{+1.2}_{-1.8})~(1.0<z\leq3.0)
\end{empheq}
in units of Gpc$^{-3}$ yr$^{-1}$ $\Delta\log L_{\nu}^{-1}$, where the subscript \lq NR\rq\ denotes non-repeating FRBs.
We note that the slopes of the luminosity functions correspond to differential number counts.
The luminosity functions at different redshift bins overlap within 1$\sigma$ uncertainties.
The luminosity functions of non-repeating FRBs do not show any significant evolution over a long range of cosmic time, i.e. $\sim$10 Gyr from $z=0$ to $z\sim1.5$.

The luminosity functions of star-forming galaxies and AGNs are known to increase by about one order of magnitude from $z\sim0$ to $z\sim1.5$ \citep{Madau2017,Hopkins2007}.
For comparison, we use the cosmic star formation-rate density, $\psi_{\rm SFRD}$, parameterised by \citet{Madau2017}:
\begin{equation}
\label{CSFRD}
\log(\psi_{\rm SFRD})=\log\left\{0.01\frac{(1+z)^{2.6}}{1+[(1+z)/3.2]^{6.2}}\right\}~(M_{\odot}~{\rm yr}^{-1}~{\rm Mpc}^{-3}),
\end{equation}
the black-hole (BH) accretion-rate density, $\psi_{\rm BHAR}$ \citep[i.e. the best-fit polynomial function of the third degree to the observed data in][]{Hopkins2007}:
\begin{equation}
\label{BHAR}
\begin{split}
\log(\psi_{\rm BHAR})&=-5.595 -5.409\times10^{-3}z +4.689\times10^{-2}z^{2} \\
 &\quad-3.341\times10^{-3}z^{3}~(M_{\odot}~{\rm yr}^{-1}~{\rm Mpc}^{-3}),
\end{split}
\end{equation}
and the cosmic stellar-mass density, $\rho_{*}$ \citep[i.e. the best-fit polynomial function of the eighth degree to the observed data in][]{Lopez2018}:
\begin{equation}
\label{CSMD}
\begin{split}
\log(\rho_{*})&=8.156+5.906\times10^{-2}z-7.111\times10^{-2}z^{2}+4.034\times10^{-2}z^{3} \\
 &\quad-1.256\times10^{-2}z^{4}+2.209\times10^{-3}z^{5}-2.216\times10^{-4}z^{6} \\
 &\quad+1.179\times10^{-5}z^{7}-2.585\times10^{-7}z^{8}~(M_{\odot}~{\rm Mpc}^{-3}).
\end{split}
\end{equation}
The cosmic stellar-mass and BH accretion-rate densities are the stellar mass per unit volume, and mass accreted onto super massive black holes per unit volume per unit time, respectively, as a function of redshift.

The redshift evolutions of $\log(\psi_{\rm SFRD})$, $\log(\psi_{\rm BHAR})$, and $\log(\rho_{*})$ between $z=0.16$ and $1.3$ are 0.7, 1.2, and $-0.3$ dex, respectively.
Here $z=0.16$ and 1.3 are the median redshifts in the low-$z$ ($0.01<z\leq0.35$) and high-$z$ ($1.0<z\leq3.0$) bins of non-repeating FRBs, respectively.
The 0.7 and 1.2 dex evolutions from the low-$z$ bin are shown by the grey shaded region in Fig. \ref{fig4}, assuming that the occurrence rate of non-repeating FRBs is proportional to star-forming or AGN activities. 
The $-0.3$ dex evolution of the cosmic stellar-mass density is shown by the grey dash-dotted line.
Our calculated luminosity function of non-repeating FRBs at $1.0<z\leq3.0$ is lower than expected if related to star-forming/AGN activities, as shown by the comparison to the grey-shaded region in Fig. \ref{fig4}a.
Therefore, the behaviour of luminosity functions of non-repeating FRBs is obviously different from those seen in star-forming galaxies and AGNs.
This indicates that non-repeating FRBs are unlikely to be related to the on-going activities of galaxies and AGNs.

\subsubsection{Repeating FRBs}
Luminosity functions of repeating FRBs are shown in Fig. \ref{fig4}b.
The data in different redshift bins show no overlap in the time-integrated luminosity, hampering comparisons of the luminosity functions at a fixed time-integrated luminosity.
We performed two different fittings: (i) a linear fit to data in each redshift bin using the slope of low-$z$ ($0.01<z\leq0.3$) data, and (ii) a linear fit to all of data presented in Fig. \ref{fig4}b.
The best-fit linear functions are shown by the coloured dashed lines for case (i) and a black dashed line for case (ii).
These functions are 
\begin{equation*}
\log\Phi_{\rm R}=
\end{equation*}
\vspace{-20pt}
\begin{empheq}[left={\empheqlbrace}]{alignat=2}
\label{LFrepeat1}
& (-1.4^{+0.5}_{-0.4}) (\log L_{\nu} -29.5)+(2.7^{+0.3}_{-0.3})(0.01<z\leq0.3)\\
\label{LFrepeat2}
& (-1.4^{+0.5}_{-0.4}) (\log L_{\nu} -29.5)+(3.4^{+1.0}_{-0.4})~(0.3<z\leq0.7)\\
\label{LFrepeat3}
& (-1.4^{+0.5}_{-0.4}) (\log L_{\nu} -29.5)+(3.9^{+1.3}_{-1.4})~(0.7<z\leq1.5)
\end{empheq}
for case (i), and 
\begin{equation*}
\log\Phi_{\rm R}=
\end{equation*}
\vspace{-20pt}
\begin{empheq}[]{alignat=2}
\label{LFrepeat4}
& (-0.97^{+0.15}_{-0.23}) (\log L_{\nu} -29.5)+(2.8^{+0.2}_{-0.2})(0.01<z\leq1.5)
\end{empheq}
for case (ii) in units of Gpc$^{-3}$ yr$^{-1}$ $\Delta\log L_{\nu}^{-1}$, where the subscript \lq R\rq\ denotes repeating FRBs.
The fitting $\chi^{2}$ (reduced $\chi^{2}$) values are 0.11 (0.11) and 1.6 (0.53) for cases (i) and (ii), respectively.
Even though case (i) indicates a small $\chi^{2}$ value, it could be due to overfitting as its reduced $\chi^{2}$ ($\ll1$), leading to no conclusive answer.
Therefore, making an assumption either on the slope or \lq no redshift evolution\rq\ is necessary in order to compare the luminosity functions.

The slope of the luminosity function of repeating FRBs at $0.01<z\leq0.3$ is demonstrated by the blue thick dashed line.
The same slopes are demonstrated by thick dashed lines through data points in green and red for $0.3<z\leq0.7$ and $0.7<z\leq1.5$, respectively.
If the slope at $0.01<z\leq0.3$ is assumed for repeating FRBs at $z>0.3$, the linear fits of luminosity functions indicate an increasing evolution towards higher redshifts.
This trend is similar to those of star-forming galaxies and AGNs.

However, due to the lack of data, it may also be possible for no redshift evolution of repeating FRBs between $z\sim0$ and 1.5.
In this case, the profile of the luminosity function can be constrained from the data.
The linear fit between $\log L_{\nu}=$ 30.4 to 31.9 is shown by the grey solid line in Fig. \ref{fig4}b.
Comparing the grey solid line to the blue solid line, the slope of the luminosity function of repeating FRBs may be slightly flatter at the brighter end.

\subsection{Volumetric occurrence rates}
\label{results_V}
The FRB volumetric occurrence rates are shown in Fig. \ref{fig5}.
We integrate the best-fit luminosity functions over $\log L_{\nu}=$31.2-32.5, and $\log L_{\nu}=$29.5-32.0 (erg Hz$^{-1}$) for non-repeating and repeating FRBs, respectively.
The upper limits of the integration are derived from the maximum time-integrated luminosities of the non-repeating and repeating FRBs selected in this work. 
While the lower bounds are arbitrarily selected, we note that altering the integration ranges does not affect the results in Section \ref{fitting} (see Section \ref{integ_range} for details). 
The slope of the luminosity function of repeating FRBs at $z>0.3$ is assumed to be the same as that at $0.01<z\leq0.3$.

In Fig. \ref{fig5}, volumetric occurrence rates of non-repeating and repeating FRBs are shown as a function of redshift, in comparison to other astronomical transients \citep{Ravi2019,Horiuchi2011,Moriya2019,Wanderman2010,Maoz2017,Nicholl2017}.
Thick solid lines indicate cosmic stellar-mass \citep{Lopez2018}, star formation-rate \citep{Madau2017}, and BH accretion-rate \citep{Hopkins2007} densities.
These cosmic densities are adjusted to the rates of non-repeating and repeating FRBs at $z=0.16$ and $z=0.17$, respectively, with the same logarithmic scale as the volumetric occurrence rate.
The values, $z=0.16$ and $z=0.17$, are median redshifts of non-repeating and repeating FRBs in the $0.01<z\leq0.35$ and $0.01<z\leq0.3$ bins, respectively.

Cosmological transients associated with star-forming activities or young stellar populations, e.g., core-collapse supernovae and long gamma-ray bursts, appear to experience increasing volumetric occurrence rates towards higher redshifts. 
This trend reflects the redshift evolution of the cosmic star formation-rate density estimated by star-forming galaxies (a blue thick line in Fig. \ref{fig5}).

\subsection{Fitting to volumetric occurrence rates}
\label{fitting}
Here we describe the fitting analysis to the volumetric occurrence rates of non-repeating and repeating FRBs in Fig. \ref{fig5}. 
We utilised the cosmic stellar-mass \citep{Lopez2018}, star formation-rate \citep{Madau2017}, and BH accretion-rate \citep{Hopkins2007} densities as \lq fiducial\rq\ fitting functions.
For the cosmic stellar-mass and BH accretion-rate densities, we derived the best-fit polynomial functions of the eighth and third degrees, respectively.
For the cosmic star formation-rate density, we used an analytic formula derived in the previous study \citep{Madau2017}. 
These functions are presented in Eqs. \ref{CSFRD}-\ref{CSMD} in Section \ref{z_evol_function}.
Constant free parameters are added to Eqs. \ref{CSFRD}-\ref{CSMD} to fit the volumetric occurrence rates of FRBs in Fig. \ref{fig5}.
The constant free parameters represent scaling factors between the FRB volumetric occurrence rates and cosmic (star formation-rate/stellar-mass/BH accretion-rate) densities in linear scale.
In order to take vertical and horizontal uncertainties in Fig. \ref{fig5} into account, Monte Carlo simulations were performed.
For data points in Fig. \ref{fig5}, we assumed Gaussian probability distribution functions with standard deviations of uncertainties of volumetric occurrence rates.
Similarly, the redshift uncertainty of each data point is also included using a median value of redshift errors of FRBs in each redshift bin.

For each simulation, we first re-sample each data point from the distributions described above, and then fit these with $\chi^2$ using the three different functions of redshift evolutions. 
Therefore, three different $\chi^2$ values corresponding to the different redshift evolutions are computed in each simulation.
The three $\chi^2$ values in each simulation are then added to the histograms in different colours in Fig. \ref{fig6}. 
In each simulation, the $\chi^{2}$ value is calculated by
\begin{equation}
\label{chi2}
\chi^{2}={\sum_{i=1}^{n}} \left[ \frac{R_{i}^{\rm FRB}(z_{i})-R_{i}^{\rm func}(z_{i})}{\sigma_{R_{i}^{\rm FRB}}} \right]^{2},
\end{equation}
where $R_{i}^{\rm FRB}$ and $R_{i}^{\rm func}$ are $i$th data of the volumetric occurrence rates of FRBs and fitting function at $z_{i}$, respectively.
$\sigma_{R_{i}^{\rm FRB}}$ is the uncertainty of $R_{i}^{\rm FRB}(z_{i})$.
These procedures were iterated 10,000 times for the randomised data points.
During the iteration, we do not reapply the redshift cut described in Section \ref{selection} to the data.
This is because altering the sample due to the redshift uncertainties does not significantly affect the derived luminosity functions and our arguments (see APPENDIX C for details).

\begin{figure*}
    \includegraphics[width=\columnwidth]{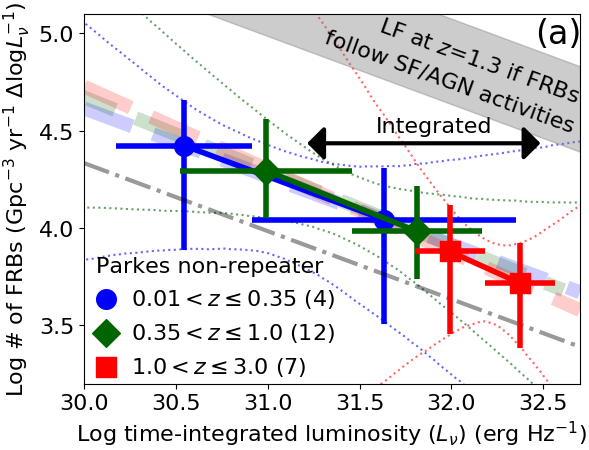}
    \includegraphics[width=\columnwidth]{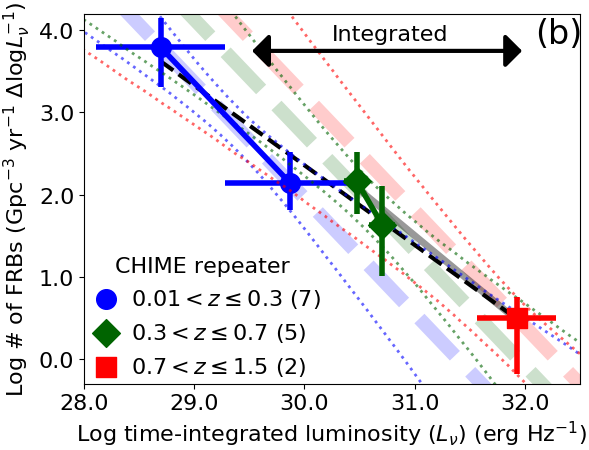}
    \caption{
    Time-integrated-luminosity functions of FRBs: the number density of FRBs per unit time-integrated luminosity in log scale per unit time, defined by $\log \Phi$ in equation (\ref{LF}), as a function of time-integrated luminosity.
    (a) Non-repeating FRBs detected by Parkes are shown by dots with errors including Poisson errors and $\delta \rho$ in the vertical axis and luminosity bins in the horizontal axis.
    Different colours correspond to different redshift bins.
    Number of sample in each redshift bin is given in parentheses in the legend.
    The linear best-fits to the data are indicated by the coloured thick dashed lines.
    The grey shaded region corresponds to 0.7-1.2 dex evolutions of the luminosity function from the low-$z$ bin ($0.01<z\leq0.35$, $z_{\rm median}=0.16$) to high-$z$ bin ($1.0<z\leq3.0$, $z_{\rm median}=1.3$), assuming that the occurrence rate of non-repeating FRBs is controlled by star-forming or AGN activities \citep{Madau2017,Hopkins2007}.
    The dash-dotted line corresponds to $-0.3$ dex evolution from the low-$z$ bin to high-$z$ bin, assuming the cosmic stellar mass-density evolution \citep{Lopez2018}.
    The dotted lines correspond to 16 and 84 percentiles (approximately $\pm$1 $\sigma$ uncertainties) of the fits derived from Monte Carlo simulations. 
    The arrow is the integrated luminosity range to calculate the volumetric occurrence rates of non-repeating FRBs in Fig. \ref{fig5}a.
    (b) Similar approach as the panel (a) except for repeating FRBs detected by CHIME.
    The repeating FRBs are counted such that the identical FRB ID indicates the same source.
    For each repeating FRB ID, the time-integrated luminosity is calculated by a median value of the individual repeated bursts.
    The coloured thick dashed lines are linear fits to data in the three redshift bins assuming a slope of the luminosity function for repeating FRBs at $0.01<z\leq0.3$.
    The black dashed line is a linear fit to all data points.
    The grey solid line is a linear fit to data at $z>0.3$.
    The dotted lines indicate fitting errors of coloured thick dashed lines including the slope uncertainty for repeating FRBs at $0.01<z\leq0.3$ and the error of each data point.
    }
    \label{fig4}
\end{figure*}

\begin{figure*}
    \includegraphics[width=2.0\columnwidth]{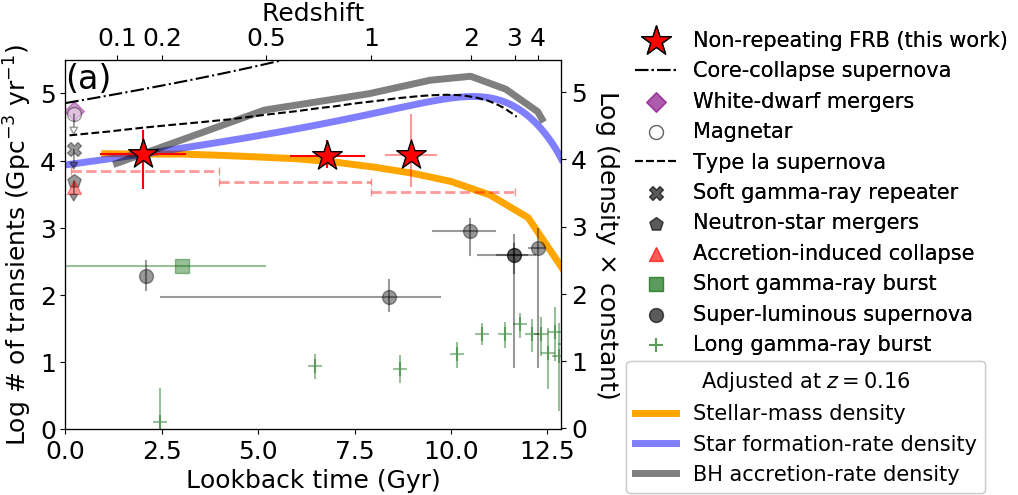}
    \includegraphics[width=2.0\columnwidth]{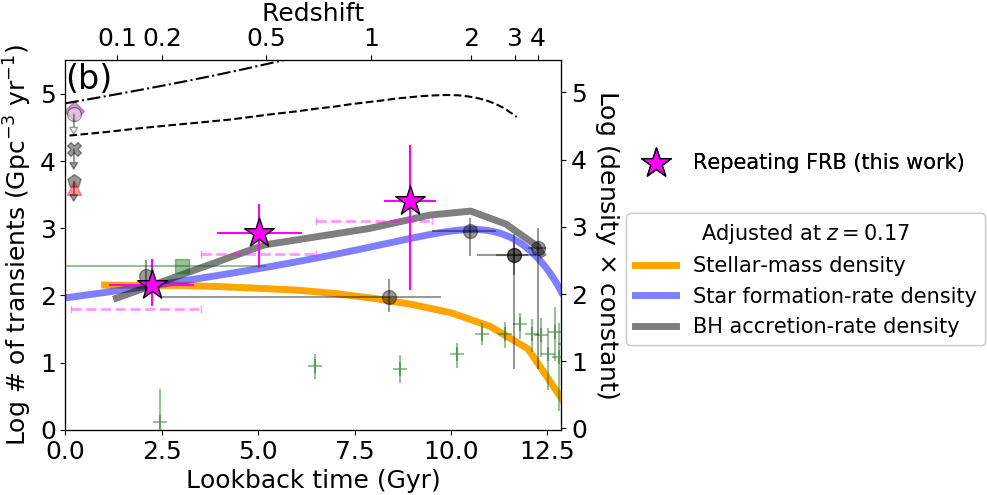}
    \caption{
    Volumetric occurrence rates of FRBs and other astronomical transients as a function of lookback time or redshift.
    (a) The occurrence rates of non-repeating FRBs detected by Parkes are shown by the red stars.
    A median redshift of non-repeating FRBs in each redshift bin is used to plot the FRB data.
    The redshift bins are shown by the red horizontal dashed lines.
    Volumetric occurrence rates of non-repeating FRBs are derived by integrating the best-fit functions to the luminosity functions in Fig. \ref{fig4}a.
    The FRBs' vertical errors include fitting uncertainties of the luminosity functions derived from Monte Carlo simulations indicated by the dotted lines in Fig. \ref{fig4}a.
    The FRBs' horizontal bars correspond to median values of redshift uncertainties of non-repeating FRBs in individual redshift bins.
    Cosmic stellar-mass \citep{Lopez2018}, star formation-rate \citep{Madau2017}, and black hole (BH) accretion-rate \citep{Hopkins2007} densities (orange, blue, and black lines, respectively) are adjusted to a volumetric occurrence rate of non-repeating FRBs at $z_{\rm median}=0.16$ with the same log scale as the volumetric occurrence rates.
    For comparison, other astronomical transients are shown by different markers \citep{Ravi2019,Horiuchi2011,Moriya2019,Wanderman2010,Maoz2017,Nicholl2017}.
    (b) Similar approach as the panel (a) except for repeating FRBs detected by CHIME.
    The repeating FRBs are counted such that the identical FRB ID indicates the same source.
    Note that the same slope is assumed for the luminosity functions at different redshifts due to the small statistics of repeating FRBs.
    }
    \label{fig5}
\end{figure*}

\begin{figure*}
    \centering
    \includegraphics[width=2.0\columnwidth]{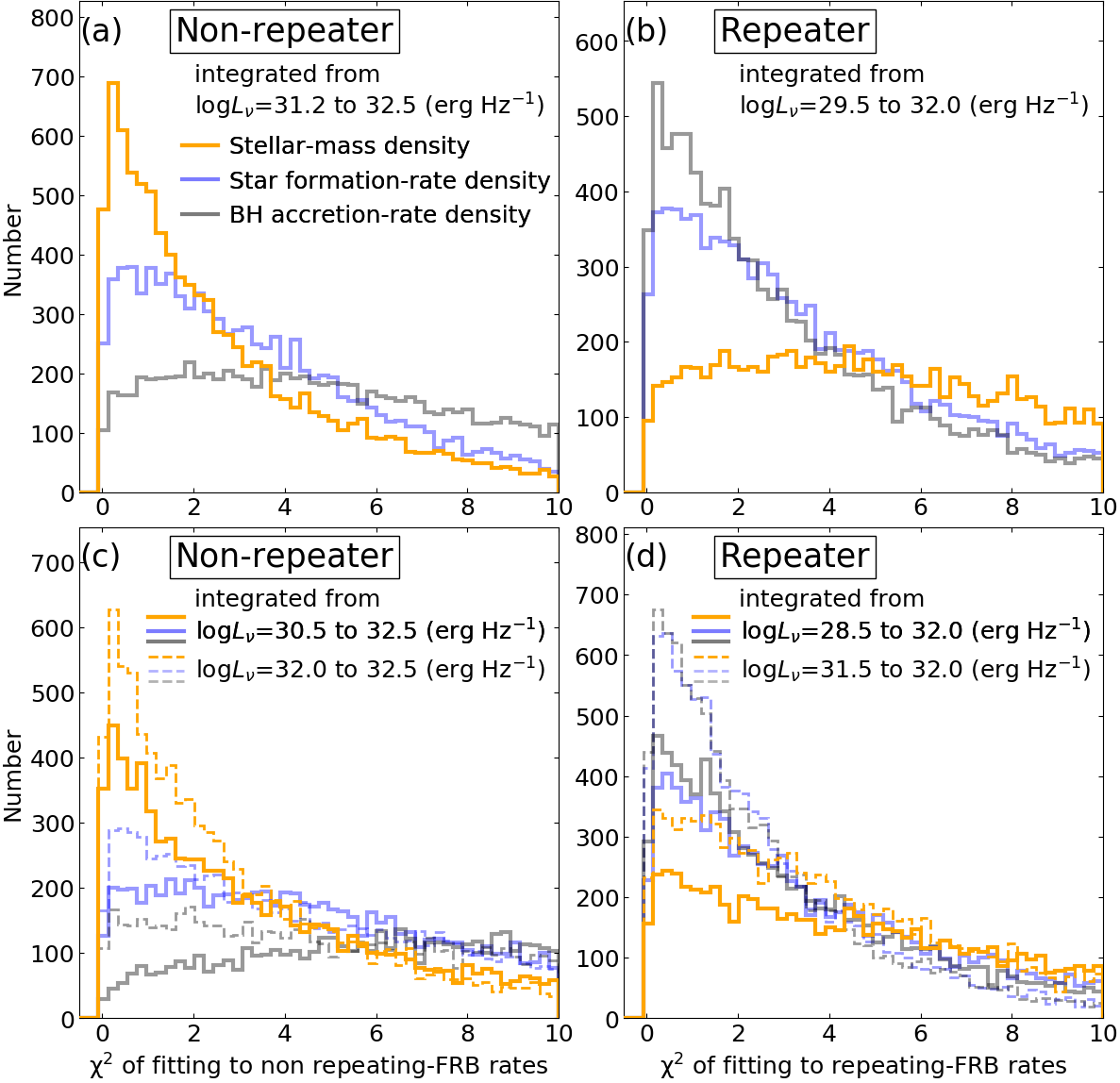}
    \caption{
    $\chi^{2}$ values of fittings to the FRB volumetric occurrence rates with the stellar-mass, star formation-rate, and black hole (BH) accretion-rate densities;
    (a) and (b) for non-repeating and repeating FRBs integrated over $\log L_{\nu}= 31.2$-$32.5$ and $\log L_{\nu}= 29.5$-$32.0$ (erg Hz$^{-1}$), respectively; 
    (c) for non-repeating FRBs integrated over $\log L_{\nu}= 30.5$-$32.5$ and $\log L_{\nu}=32.0$-$32.5$ (erg Hz$^{-1}$) (solid and dashed histograms, respectively);
    (d) for repeating FRBs integrated over $\log L_{\nu}= 28.5$-$32.0$ and $\log L_{\nu}=31.5$-$32.0$ (erg Hz$^{-1}$) (solid and dashed histograms, respectively).
    The fitting procedure is iterated 10,000 times with randomly assigned errors on volumetric occurrence rates and redshifts in Fig. \ref{fig5}.
    The assigned random errors follow Gaussian probability distributions with standard deviations of the measured errors.
    }
    \label{fig6}
\end{figure*}

In summary, we performed the following procedures in Section \ref{results}.
\begin{description}
\item[(1) fit a linear function to data in each redshift bin in Fig \ref{fig4}.]\mbox{}\\
For non-repeating FRBs, $\Phi(z_{\rm median}, L_{\nu})$ is derived for each redshift bin (Eqs. \ref{LFnorepeat1}-\ref{LFnorepeat3}). 
For repeating FRBs, we assume either (i) $\Phi(z_{\rm median},L_{\nu})=f(z_{\rm median})\Phi(z_{\rm median}=0.17, L_{\nu})$ (Eqs. \ref{LFrepeat1}-\ref{LFrepeat3}) or (ii) $\Phi(z_{\rm median},L_{\nu})=\Phi(L_{\nu})$ without redshift evolution (Eq. \ref{LFrepeat4}) due to their small number statistics, where $f(z_{\rm median})$ represents a factor of redshift evolution.\\
\item[(2) integrate $\Phi(z_{\rm median}, L_{\nu})$ over $L_{\nu}$.]\mbox{}\\
For each population, we integrate $\Phi(z_{\rm median}, L_{\nu})$ over the ranges shown in Fig. \ref{fig4} to derive data points in Fig. \ref{fig5}.
For repeating FRBs, we assume case (i), i.e., Eqs. \ref{LFrepeat1}-\ref{LFrepeat3}.\\
\item[(3) fit Eqs. \ref{CSFRD}-\ref{CSMD} to the data points in Fig. \ref{fig5}.]\mbox{}\\
A constant free parameter is added to each of Eqs. \ref{CSFRD}-\ref{CSMD} for the fitting.
In order to take vertical and horizontal uncertainties of data points in Fig. \ref{fig5} into account, we performed Monte Carlo simulations by randomising data points following Gaussian probability distributions. 
\end{description}

Fig. \ref{fig6} shows histograms of $\chi^{2}$ calculated by the Monte Carlo simulations.
Figs. \ref{fig6}a and \ref{fig6}b are fits to the occurrence rates of non-repeating and repeating FRBs, respectively.
For the non-repeating FRBs, $\chi^{2}$ of the cosmic stellar-mass density shows a sharp peak at $\chi^{2}\sim0.5$ compared to other cosmic densities ($\chi^{2}\sim1$ and 4 at the peaks of histograms of star formation-rate and BH accretion-rate densities, respectively).
The histograms clearly indicate that the cosmic stellar-mass density is the best match with the volumetric occurrence rates of non-repeating FRBs.
For repeating FRBs, the cosmic star formation-rate and BH accretion-rate densities show more peaky histograms at $\chi^{2}\sim0.5$ than the cosmic stellar-mass density.
The repeating FRBs show better fits with the cosmic star formation-rate and BH accretion-rate densities.
We note that the slope of luminosity function of repeating FRBs at $0.01<z\leq0.3$ is assumed for that at $z>0.3$.
The fitting result would strongly depend on this assumption.
Future observational constraints on the slopes are necessary to obtain a conclusive answer to this point.

\section{Discussion}
\label{discussion}
\subsection{Systematic uncertainties}
\label{systematics}
\subsubsection{Unknown locations of FRBs in field of view}
We discuss systematic uncertainties of the FRB luminosity functions and volumetric occurrence rates.
Since the exact locations of FRBs in the fields of view of Parkes and CHIME are unknown, accurate sensitivity corrections for individual FRBs are difficult.
Therefore, fluences of Parkes and CHIME FRBs reported in FRBCAT are in principle lower limits.
The actual fluence can vary by a factor of 1.7 on average, depending on the slope of the source count distribution of FRBs \citep{Macquart2018}.
The accurate measurement of the slope is still in debate \citep[e.g.,][]{James2019}.
The luminosity functions could shift towards the brighter end by $\log$(1.7)$\sim$0.2 dex on average due to this systematic uncertainty.
The volumetric occurrence rates integrated over the luminosity functions are not affected by this uncertainty, since the number densities of luminosity functions do not change.

\subsubsection{Source count distribution}
The source count distribution of FRBs also affects the effective survey area.
The effective survey area can increase by a factor of $\sim$3 depending on the slope of the source counts of FRBs \citep{Macquart2018}.
This uncertainty could systematically change the number densities of luminosity functions.
Therefore the volumetric occurrence rates in this work have the systematic uncertainty of at least $\log(3)\sim0.5$ dex.
A major focus in this work is the relative differences in the volumetric occurrence rates from $z\sim0$ to $\sim$1.5.
This systematic uncertainty is offset when the volumetric occurrence rates are compared to the cosmic densities, i.e., stellar-mass, star formation-rate, and BH accretion-rate densities.
Therefore this uncertainty does not affect our arguments significantly.

\subsubsection{Detection threshold}
The uncertainty on calculating the detection threshold, $E_{\rm lim}$, is also a potential systematic error.
Since we empirically derived the thresholds based on the current FRB sample, it is possible that the thresholds may change with better statistics of future FRB data.
To check how this uncertainty may affect the volumetric occurrence rates, we tested two cases of $E_{\rm lim1}=0.8E_{\rm lim}$ and $E_{\rm lim2}=1.2E_{\rm lim}$ for the Parkes non-repeating FRBs, where $E_{\rm lim}=0.72w_{\rm obs}^{1/2}$ (Jy ms).
We found that the differences in occurrence rates from the fiducial case with $E_{\rm lim}=0.72w_{\rm obs}^{1/2}$ are at most 0.007, 0.05, and 0.2 dex at $0.01<z\leq0.35$, $0.35<z\leq1.0$, and $1.0<z\leq3.0$, respectively.
These values are much smaller than the uncertainties on individual data points in Fig. \ref{fig5}a.
Therefore we conclude that, at least, the 20\% uncertainty on calculating $E_{\rm lim}$ does not affect our arguments significantly.
Even if different source count distributions are assumed, the detected number of FRBs at each redshift bin does not change statistically, since (i) the empirically-defined $E_{\rm lim}$ varies linearly with fluence, e.g., if the fluence is actually 1.7 times higher on average, then $E_{\rm lim}$ is also 1.7 times higher and (ii) the dispersion measure-derived redshift is independent of the source counts. 
The number of non-repeating FRBs at $1.0<z\leq3.0$ (red line in Fig. \ref{fig4}a) is about one order of magnitude smaller than the expected one from the cosmic star formation-rate density (grey shaded region in Fig. \ref{fig4}a). 
There is no correlation with either the assumed source count distribution or assumed fluences/survey areas. 

\subsubsection{Integration range of luminosity function}
\label{integ_range}
The absolute values of volumetric occurrence rates shown in Fig. \ref{fig5} depend on the integration ranges of luminosity functions.
Our main focus in this work is relative differences of FRB occurrence rates at different redshifts.
The different integration ranges of luminosity functions may result in different dependencies of FRB rates on redshift.
To investigate this uncertainty, we integrated the luminosity functions between $\log L_{\nu}=$ 30.5 and 32.5 (erg Hz$^{-1}$) for non-repeating FRBs and 28.5 and 32.0 (erg Hz$^{-1}$) for repeating FRBs, which cover the entire range of data points in Fig. \ref{fig4}.
In addition, we consider integration ranges over $\log L_{\nu}=32.0$-$32.5$ and $31.5$-$32.0$ (erg Hz$^{-1}$) for non-repeating and repeating FRBs, respectively.
These cover the $L_{\nu}$ ranges of high-$z$ data points in Fig. \ref{fig4}, where no extensive extrapolation is necessary for high-$z$ luminosity functions (i.e. red dashed lines in Figs. \ref{fig4}a and b).
The same analyses described in Section \ref{results_V} and \ref{fitting} are repeated for these new integration ranges.

The $\chi^{2}$ values of fittings to the occurrence rates of non-repeating and repeating FRBs are shown in Figs. \ref{fig6}c and \ref{fig6}d, respectively.
In Fig. \ref{fig6}c, the cosmic stellar-mass density shows much sharper histograms (dashed and solid orange histograms) peaking at $\chi^{2}\lesssim0.5$, compared to those of the cosmic star formation-rate and BH accretion-rate densities.
This indicates better fits of the cosmic stellar-mass density to non-repeating FRBs.
Similarly, in Fig. \ref{fig6}d, the star formation-rate and BH accretion-rate densities show better fits to the occurrence rates of repeating FRBs with peaks at $\chi^{2}\sim0.5$.
Therefore, the different integration ranges do not seem to change our arguments.

\subsubsection{Spectral shape}
We use the characteristic spectral index of FRBs, $\alpha=-1.5$ \citep{Macquart2019}, if the individual measurement is not available.
The characteristic broadband spectral index might be much flatter than this value \citep{Sokolowski2018,Chawla2017}.
The strongest constraint provided by multi-band observations is $\alpha\gtrsim-0.3$ \citep{Chawla2017}. 
Comparing $\alpha=-1.5$ and $-0.3$ cases, the $K$-correction term in equation (\ref{eqLnu}), i.e., $(\nu_{\rm rest}/\nu_{\rm obs})^{\alpha}/(1+z)^{2+\alpha}$, becomes $\sim2$ times smaller for $\alpha=-0.3$ than that for $\alpha=-1.5$ at $z=1.5$. 
This means that the luminosity function of non-repeating FRBs at $1.0<z\leq3.0$ might shift towards fainter luminosities by $\log(2)\sim0.3$ dex, while the luminosity functions at lower redshifts are less affected by the $K$-correction. 
This uncertainty will enhance the difference between the derived luminosity function at $1.0<z\leq3.0$ (red in Fig. \ref{fig4}a) and the expected one from the cosmic star formation-rate density (grey shaded region in Fig. \ref{fig4}a).

Many FRBs appear band-limited, i.e., their spectra are confined in narrow frequency ranges rather than that of broad-band spectra. 
This makes comparing identical frequency at source frame difficult due to the wide redshift range of our sample (Fig. \ref{fig2}).
Therefore, bolometric energies of FRBs integrated over a particular frequency range would be a better indicator for the luminosity function, i.e., bolometric-energy function.
In this case, the $K$-correction term is $1/(1+z)$ which arises from the time dimension of fluence.
For Parkes non-repeating FRBs, the characteristic $K$-correction terms (band-limited case) are $\log [1/(1+z_{\rm median})]=-0.06$ and $-0.36$ dex for low-$z$ ($0.01<z\leq0.35$) and high-$z$ ($1.0<z\leq3.0$) bins, respectively.
The characteristic $K$-correction terms calculated in Section \ref{calcLnu} (broad-band case) are $\log[(\nu_{\rm rest}/\nu_{\rm obs})^{\alpha}/(1+z_{\rm median})^{2+\alpha}]=-0.23$ and $-0.38$ dex for low-$z$ and high-$z$ bins, respectively.
Therefore, the difference of $K$-correction terms between low-$z$ and high-$z$ bins is $-0.3$ dex for the band-limited case and $-0.15$ dex for the broad-band case.
This indicates that in the band-limited case, a systematically larger $K$-correction towards lower bolometric energies is necessary for the high-$z$ non-repeating FRBs compared to the broad-band case.

When we apply the $K$-correction term, the difference becomes larger between the observed bolometric-energy function (of non-repeating FRBs at $1.0<z\leq3.0$) and the expected one from the cosmic star formation-rate density.
Hence, this supports that the cosmic stellar mass-density evolution shows a better fit to non-repeating FRBs than the cosmic star formation-rate density (see Section \ref{results} for details).
Similarly, the difference between luminosity functions of CHIME repeating FRBs at low-$z$ ($0.01<z\leq0.3$) and high-$z$ ($0.7<z\leq1.5$) bins is about one order of magnitude in the time-integrated luminosity (Fig. \ref{fig4}b).
This difference will decrease by $\sim0.15$ dex in the band-limited case, which is much smaller than the one order of magnitude difference.
Therefore, the uncertainties on the broad-band spectral index and band-limited case do not change our conclusion.

\subsubsection{Distinction between non-repeaters and repeaters}
Non-repeating and repeating FRBs are observationally defined based on their repetitions.
Therefore, it is still uncertain whether non-repeaters are definitely one-offs.
On average, the repeating FRBs are fainter than non-repeating FRBs \citep[e.g., ][]{Hashimoto2020}.
\cite{Kumar2019} reported repeating bursts from FRB 171019 which was previously thought to be a non-repeating FRB.
Its repeating bursts are $\sim$590 times fainter than the first burst discovered by ASKAP.
The observed number fraction of repeaters to non-repeaters should depend on the duration of observation \citep[e.g., ][]{Ai2020}, since the probability to find repeaters becomes higher for a longer exposure.
Therefore, it is possible that some of the non-repeating FRBs are actually repeating, but only the luminous bursts (or bursts coincident with observations) were detected due to sensitivity limits of telescopes (or the time limits of observations).

There will be a clear aversion to discovering repeating FRBs in the high-$z$ Universe, since repeaters are typically fainter than non-repeaters, and their observed duration would correspond to a shorter source-frame duration at higher redshifts. 
Correcting this effect would shift detections of high-$z$ non-repeating FRBs ($1.0<z\leq3.0$) in Fig. \ref{fig4}a to high-$z$ repeating FRBs ($0.7<z\leq1.5$) in Fig. \ref{fig4}b. 
Thus, the luminosity function of non-repeating FRBs will decrease at higher redshifts, while that of repeating FRBs will increase. 
This systematic adds even more credence to the preference of the cosmic stellar mass-density evolution for non-repeating FRBs and the cosmic star formation-rate evolution for repeating FRBs. 
The steep slope of the luminosity function of repeating FRBs (Fig. \ref{fig4}b) could be partly attributed to this systematic effect.

\subsubsection{Redshift cut and bins}
\label{zcut}
The redshift cut in Section \ref{selection} is applied before simulating the redshift uncertainty described in Section \ref{calcz}.
Therefore, low-z FRBs including non-repeating FRB 110214 at $z=0.04\pm0.07$ and repeating FRB 180908.J1232+74 at $z=0.07\pm0.07$ sometimes go out of the redshift cut within the uncertainties.
This uncertainty has negligible effects on the luminosity functions in Section \ref{results} and our conclusions in Section \ref{conclusion} (see APPENDIX C for details).

We selected redshift bins with equal time intervals, corresponding to approximately 4 and 3 Gyr for non-repeating and repeating FRBs, respectively (Section \ref{calcLF}). 
To examine the sensitivity of our results to the chosen redshift bin edges, we tested two additional cases: (i) changing the redshift bins while keeping the sample size the same in each bin, and (ii) optimising the redshift bins so that the mean value of $V_{4\pi}/V_{\rm max,4\pi}$ in each bin approaches 0.5 \citep[e.g.,][]{Avni1980,Jurek2013}, where $V_{4\pi} = 4\pi/3(d_{\rm FRB}^{3}-d_{\rm min}^{3})$ and $d_{\rm FRB}$ is the comoving distance to each FRB.

Based on these tests, we repeated the same analyses as described in Sections \ref{analysis} and \ref{results}. 
As a result, we found no significant difference in terms of the luminosity functions and $\chi^{2}$ distributions (see APPENDIX C for details).

The systematic uncertainties discussed in Section \ref{systematics} are not explicitly indicated in either Fig. \ref{fig4} or Fig. \ref{fig5} for clarity purpose, as they do not affect our arguments significantly.

\subsection{Luminosity function}
\citet{Luo2020} derived the luminosity function using a total of 46 FRBs detected with Parkes, Arecibo, Green Bank Telescope, UTMOST, and ASKAP.
Their sample includes two repeating FRBs (FRB 121102 and 171019) and 44 non-repeating ones.
They used flux densities to calculate isotropic luminosities integrated over the frequency in units of erg s$^{-1}$.
Here we use the fluence which is less affected by the time resolution of the observation and is thus preferable over the flux density \citep{Macquart2018}. 
In this work, homogeneous samples are used to avoid systematic uncertainties between different telescopes, i.e., 23 Parkes non-repeating FRBs and 14 CHIME repeating FRBs.
We derived the luminosity functions of non-repeating and repeating FRBs separately.
This is because these two populations show different observational properties such as duty factors ($\equiv <S>^{2}/<S^{2}>$ where $S$ is the flux density), rotation measures, durations, luminosities, and luminosity functions \citep[e.g., ][]{Katz2019,CHIME8repeat2019,Hashimoto2020,Fonseca2020}.
Furthermore, \citet{Luo2020} assumed a Schechter luminosity function without any redshift evolution, while the $V_{\rm max}$ method in this work does not make any assumption on the profile nor the evolution of the luminosity function across redshifts. 

Since the sample in \citet{Luo2020} is mostly dominated by non-repeating FRBs, we compare their results to ours for non-repeating FRBs only.
They reported a luminosity function with a power-law slope of $\alpha_{\rm LF}=-0.79^{+0.31}_{-0.35}$ (slope in logarithmic scale).
This value is consistent with the slopes of non-repeating FRBs in this work, i.e., $\alpha_{\rm LF}=-0.35^{+0.51}_{-0.50}$ ($0.01<z\leq0.35$), $-0.38^{+0.37}_{-0.50}$ ($0.35<z\leq1.0$), and $-0.43^{+1.19}_{-1.05}$ ($1.0<z\leq3.0$), within their 1-$\sigma$ uncertainties.

\subsection{Volumetric occurrence rates as a function of redshift}
\subsubsection{Non-repeating FRBs}
\citet{Cao2018} constrained the local volumetric occurrence rate using 22 Parkes FRBs.
The constrained value is $\sim$(3-6)$\times10^{4}$ Gpc$^{-3}$ yr$^{-1}$ for an adopted minimum FRB energy of $3\times10^{39}$ erg.
This minimum energy corresponds to $\log L_{\nu}=30.9$ erg Hz$^{-1}$, assuming the Parkes bandwidth of 0.34 GHz \citep{Cao2018}.
For comparison, we integrated the luminosity function of Parkes non-repeating FRBs at $0.01<z\leq0.35$ (Eq. \ref{LFnorepeat1}) over $\log L_{\nu}=30.9$ to 33.0 erg Hz$^{-1}$.
The derived volumetric occurrence rate is $(2.2^{+2.8}_{-1.5})\times10^{4}$ Gpc$^{-3}$ yr$^{-1}$. 
This value is consistent with \citet{Cao2018} within its 1-$\sigma$ uncertainty, as well as \citet{Deng2019}'s estimate of $(3.2\pm0.3)\times10^{4}$ Gpc$^{-3}$ yr$^{-1}$.

\citet{Cao2018} scaled the FRB rate with the cosmic star formation-rate density with some delay times and demonstrated that this can fit to the redshift and energy distributions of 22 Parkes FRBs.
In their analysis, the profiles of the energy functions (corresponding to the luminosity functions in this work) and delay times are parameterised for the fitting, although these two parameters are moderately degenerate \citep[c.f. Fig. 4 in][]{Cao2018}.
Their analysis is confined to models assuming the cosmic star formation-rate density, without testing for the cosmic stellar-mass density.
In this work, there is no degeneracy between the derivations of our luminosity functions and their redshift evolution for Parkes non-repeating FRBs, because the luminosity functions are derived at different redshift bins independently.
We note that, in this work, the analyses are quantified by taking both the star formation-rate density and the stellar-mass density into account.

In contrast to core-collapse supernovae and long gamma-ray bursts, the occurrence rate of non-repeating FRBs shows no significant increase towards higher redshifts and is almost constant (Fig. \ref{fig5}a).
The independence of the volumetric occurrence rate on redshift does not match the rising trends of the cosmic star formation-rate density and the BH accretion-rate density at $z\sim1$-2.
The volumetric occurrence rates of non-repeating FRBs are consistent with the cosmic stellar-mass density that shows an almost flat trend towards higher redshifts up to $z\sim1.5$ (orange solid line in Fig. \ref{fig5}a, see also Figs. \ref{fig6}a and \ref{fig6}c for quantitative analysis).
Therefore, the amount of stars in the Universe accumulated via past star formations is probably controlling the occurrence rate of non-repeating FRBs, instead of any on-going galactic activity.
Since old stellar populations dominate the stellar mass in the Universe \citep[e.g.,][]{Chabrier2003}, non-repeating FRBs are likely to originate from old stellar populations or long-living objects (order of Gyr).
In this sense, FRB models assuming only older populations, e.g., white dwarfs \citep[e.g.,][]{Li2018}, neutron stars \citep[e.g.,][]{Yamasaki2018}, and BHs \citep[e.g.,][]{Liu2016}, would be favoured for a majority of non-repeating FRBs up to $z\sim1.5$.
Our findings disfavour models involving young stellar populations or their remnants with shorter lifetimes (order of $\lesssim$\,Myr), e.g., supernova remnants \citep[e.g.,][]{Murase2016}, magnetars \citep[e.g.,][]{Metzger2019,Lu2020}, pulsars \citep[e.g.,][]{Katz2017a}, and AGNs \citep[e.g.,][]{Gupta2018}, as progenitors of non-repeating FRBs.

If the occurrence rates of non-repeating FRBs are controlled by stellar mass, the probability of finding non-repeating FRBs should be higher in galaxies with stellar masses of $\log M_{*}=$10-11 ($M_{\odot}$), since such galaxies mostly contribute to the cosmic stellar-mass density up to $z\sim2$ \citep[e.g.,][]{Davidzon2017}.
Recently, host galaxies of non-repeating FRBs were identified for FRB 180924 \citep{Bannister2019}, 181112 \citep{Prochaska2019}, and 190523 \citep{Ravi2019host}, from which two (FRB 180924 and 190523 hosts) are massive galaxies with stellar masses of $\log M_{*}>10.0$ $(M_{\odot})$, and one (FRB 181112 host) has a moderate stellar mass of $\log M_{*}\sim9.5$ ($M_{\odot}$).
In addition, \citet{Macquart2020} reported four new host identifications for FRB 190102, 190608, 190611, and 190711.
Among the four host galaxies, stellar masses of two host galaxies of non-repeating FRBs (190102 and 190608) are reported by \citet{Bhandari2020}.
Their stellar masses are $\log M_{*}=9.5$ and $10.4$ $M_{\odot}$, respectively.
\citet{Bhandari2020} argued that the host galaxies of ASKAP non-repeating FRBs exhibit lower star formation relative to their high stellar mass, while they are not being confined in a well-defined locus of a particular class.
Thus, these observational results of non-repeating FRB host galaxies are consistent with our argument.

\subsubsection{Repeating FRBs}
\citet{James2019b} estimated an upper limit on the volumetric density of repeating FRBs at $z\sim0$ ($<$27 Gpc$^{-3}$) using ASKAP FRBs with fluences between its detection threshold and a burst energy cut-off at $10^{42}$ erg.
The corresponding time-integrated-luminosity range is $\log L_{\nu} = \sim 30.5$ to $\sim 33$ erg Hz$^{-1}$ \citep{Hashimoto2020} assuming 1 GHz emission bandwidth \citep{James2019b}.
We integrated the luminosity function of CHIME repeating FRBs at $0.01<z\leq0.3$ between these time-integrated luminosities.
The derived volumetric occurrence rate is $6.2^{+29.9}_{-4.9}$ Gpc$^{-3}$ yr$^{-1}$. 
The observed duration of 399 days for the CHIME repeating FRBs selected in this work corresponds to $399/(1+0.17) = 341$ days at the source frame, where 0.17 is the median redshift of CHIME repeating FRBs at $0.01<z\leq0.3$. 
Therefore, the volumetric density of CHIME repeating FRBs is estimated to be $(6.2^{+29.9}_{-4.9})\times341/365 = 5.8^{+27.9}_{-4.6}$ Gpc$^{-3}$ under this $L_{\nu}$ integration range. 
This value is consistent with the upper limit of $<$27 Gpc$^{-3}$ derived by \citet{James2019b}.

The redshift evolution of the volumetric occurrence rates of repeating FRBs is more uncertain than that of non-repeating FRBs due to the limited sample.
If the slopes of luminosity functions of repeating FRBs at $z>0.3$ are similar to that at $0.01<z\leq0.3$ (Fig. \ref{fig4}b), the volumetric occurrence rate increases towards higher redshifts (Fig. \ref{fig5}b). 
This may favour young stellar populations or AGN activities as origins of repeating FRBs (see also Figs. \ref{fig6}b and \ref{fig6}d for quantitative analysis).
A star-forming dwarf galaxy, whose star-formation rate and stellar mass are SFR=0.4 $M_{\odot}$ yr$^{-1}$ and $\log M_{*}=$7.6-7.8 $M_{\odot}$, respectively, is found to host a repeating FRB 121102 \citep{Tendulkar2017}.
FRB 180916.J0158+65 was localised at a star-forming region in a nearby spiral galaxy \citep{Marcote2020}.
Recently FRB 200428 was found to be spatially coincident with the known Galactic soft gamma-ray repeater 1935+2154 \citep{Scholz2020,Bochenek2020}.
It shows multiple episodes of gamma-ray outbursts \citep{Lin2020} and repeating radio bursts \citep[e.g.,][]{Kirsten2020}.
This source is identified as a magnetar via subsequent X-ray follow-up observations \citep{Israel2016}, spatially associated with the supernova remnant G57.2+0.8 \citep[e.g.,][]{Surnis2016}.
These observational results support the hypothesis that repeating FRBs trace young stellar populations or their remnants with $\lesssim$\,Myr time scales.

However, we caution that it is also possible for the occurrence rate of repeating FRBs not to change with redshift or decline at the high-redshift Universe, because there is no constraint on the slopes of their luminosity functions at $z>0.3$ (Fig. \ref{fig4}b).
If this is the case, the slope of luminosity function of repeating FRBs might be flatter at the bright end compared to the faint end.
More future data is necessary to conclude this point.

The future Square Kilometre Array (SKA) \citep{Dewdney2009} is one of the most ideal instruments to address the origins of non-repeating and repeating FRBs.
A massive amount of high-redshift FRBs will be detected by the SKA \citep[e.g., ][]{Fialkov2017, Hashimoto2020c}, allowing us to accurately constrain the luminosity functions and volumetric occurrence rates.
The high-spatial resolution of the SKA will also significantly increase the number of FRBs localised down to $\sim$arcsecond level on the sky.
In the SKA era, the host galaxies can be increasingly identified by follow-up observations with optical to near-infrared telescopes, providing a more direct evidence of the stellar populations of FRB host galaxies and thus indications of progenitors.

\section{Conclusions}
\label{conclusion}
Based on the FRB catalogue as of 24 Feb. 2020, we presented the luminosity functions and volumetric occurrence rates of non-repeating and repeating FRBs as a function of redshift.
The luminosity functions and volumetric occurrence rates of non-repeating FRBs do not show redshift evolution from $z=0$ to $\sim1.5$ (Figs. \ref{fig4}a and \ref{fig5}a), in contrast to other cosmological transients such as supernovae and gamma-ray bursts.
The nearly-constant rate over the past $\sim$10 Gyr is consistent with the flat trend of cosmic stellar-mass density up to $z\sim1.5$ traced by old stellar populations (Figs. \ref{fig6}a and \ref{fig6}c).
This indicates that the occurrence rate of non-repeating FRBs could be controlled by the stellar-mass evolution of long-living objects with $\sim$Gyr time scales.
White dwarfs, neutron stars, and black holes are favoured as likely progenitors of non-repeating FRBs.

In contrast, the occurrence rates of repeating FRBs might increase towards higher redshifts up to $z\sim1.5$ (Figs. \ref{fig4}b and \ref{fig5}b) if the slope of their luminosity function does not evolve as a function of redshift.
This increasing trend shows better fits with the cosmic star formation-rate and black hole accretion-rate densities (Figs. \ref{fig6}b and \ref{fig6}d).
This suggests that repeating FRBs could be associated with young stellar populations (or their remnants) or AGN activity with $\lesssim$ Myr time scales, favouring e.g. supernova remnants, young pulsars, magnetars, and AGNs as their progenitor candidates.
We caution that it is also possible for the occurrence rate of repeating FRBs not to change with redshift or decline at the high-redshift Universe, since there is no constraint on the slopes of their luminosity functions
at $z > 0.3$ (Fig. \ref{fig4}b).
More future data is necessary to conclude the possible redshift evolution of the volumetric occurrence rates of repeating FRBs.

\section*{Acknowledgements}
We are very grateful to the anonymous referee for many insightful comments.
TH and AYLO are supported by the Centre for Informatics and Computation in Astronomy (CICA) at National Tsing Hua University (NTHU) through a grant from the Ministry of Education of the Republic of China (Taiwan).
TG acknowledges the support by the Ministry of Science and Technology of Taiwan through grant 108-2628-M-007-004-MY3.
AYLO's visit to NTHU was supported by the Ministry of Science and Technology of the ROC (Taiwan) grant 105-2119-M-007-028-MY3, hosted by Prof. Albert Kong.
This work used high-performance computing facilities operated by the CICA at NTHU. 
This equipment was funded by the Ministry of Education of Taiwan, the Ministry of Science and Technology of Taiwan, and NTHU.
This research has made use of NASA's Astrophysics Data System.

\section*{Data availability}
The data underlying this article are available in the FRBCAT project at \url{http://frbcat.org/} and references therein.
The full catalogue of derived physical parameters of FRBs are available at \url{http://www.phys.nthu.edu.tw/~tetsuya/}.


\bibliographystyle{mnras}
\bibliography{FRB_LFz_mnras} 



\appendix
\section{FRB catalogue}
Observed and derived parameters of FRBs selected in Section \ref{selection} are summarised in Tables \ref{tabA1} and \ref{tabA2}, respectively.

\begin{table*}
	\centering
	\caption{
	Observed parameters of FRBs used for the calculations of their luminosity functions.
    }
	\label{tabA1}
	\begin{flushleft}
	\begin{tabular}{|l|c|c|c|c|c|c|c|c|c|}\hline
\multicolumn{10}{|c|}{Non-repeating FRB} \\
(1)& (2)            & (3)       & (4)             & (5)                   & (6)        & (7)              & (8)  & (9) & (10)\\ 
ID  & $l$ & $b$ & DM & $E_{\nu_{\rm obs}}$ & $w_{\rm obs}$ & $\alpha$ & $\nu_{\rm obs}$ & $\Delta \nu_{\rm obs}$ & $w_{\rm sample}$ \\
 & (deg) & (deg) & (pc cm$^{-3}$)  & (Jy ms)  & (ms) &          & (GHz)           &     (MHz)     &  (ms) \\ \hline 
010125 & 356.64 & -20.02 & 790.00$\pm$3.00 & 2.82$^{+2.99}_{-2.27}$ & 9.40$\pm$2.65 & $-1.50$ & 1.37 & 3.00 & 0.12 \\
010312 & 274.72 & -33.30 & 1187.00$\pm$14.00 & 6.08$^{+0.33}_{-0.33}$ & 24.30$\pm$12.50 & $-1.50$ & 1.37 & 3.00 & 1.00 \\
010621 & 25.43 & -4.00 & 745.00$\pm$10.00 & 2.87$^{+5.24}_{-2.11}$ & 7.00$\pm$3.12 & $-1.50$ & 1.37 & 3.00 & 0.25 \\
010724 & 300.65 & -41.81 & 375.00$\pm$3.00 & 150.00$^{+4.69}_{-5.33}$ & 5.00$\pm$8.00 & $-1.50$ & 1.37 & 3.00 & 1.00 \\
090625 & 226.44 & -60.03 & 899.55$\pm$0.01 & 2.19$^{+2.10}_{-1.44}$ & 1.92$\pm$0.80 & $-1.50$ & 1.35 & 0.39 & 0.06 \\
110214 & 290.70 & -66.60 & 168.90$\pm$0.50 & 51.30$^{+2902.70}_{-24.30}$ & 1.90$\pm$0.90 & $-1.50$ & 1.35 & 0.39 & 0.06 \\
110220 & 50.83 & -54.77 & 944.38$\pm$0.05 & 7.28$^{+0.13}_{-0.13}$ & 5.60$\pm$0.10 & $-1.50$ & 1.35 & 0.39 & 0.06 \\
110703 & 81.00 & -59.02 & 1103.60$\pm$0.70 & 2.15$^{+2.73}_{-1.41}$ & 4.30$\pm$2.04 & $-1.50$ & 1.35 & 0.39 & 0.06 \\
121002 & 308.22 & -26.26 & 1629.18$\pm$0.02 & 2.34$^{+4.46}_{-0.91}$ & 5.44$\pm$2.35 & $-1.50$ & 1.35 & 0.39 & 0.06 \\
130626 & 7.45 & 27.42 & 952.40$\pm$0.10 & 1.47$^{+2.45}_{-0.59}$ & 1.98$\pm$0.83 & $-1.50$ & 1.35 & 0.39 & 0.06 \\
130628 & 225.96 & 30.66 & 469.88$\pm$0.01 & 1.22$^{+0.47}_{-0.43}$ & 0.64$\pm$0.13 & $-1.50$ & 1.35 & 0.39 & 0.06 \\
130729 & 324.79 & 54.74 & 861.00$\pm$2.00 & 3.43$^{+6.55}_{-2.39}$ & 15.61$\pm$8.12 & $-1.50$ & 1.35 & 0.39 & 0.06 \\
131104 & 260.55 & -21.93 & 779.00$\pm$1.00 & 2.33$^{+2.17}_{-0.88}$ & 2.08$\pm$0.67 & $-1.50$ & 1.35 & 0.39 & 0.06 \\
140514 & 50.84 & -54.61 & 562.70$\pm$0.60 & 1.32$^{+0.64}_{-1.66}$ & 2.80$\pm$1.38 & $-1.50$ & 1.35 & 0.39 & 0.06 \\
150215 & 24.66 & 5.28 & 1105.60$\pm$0.80 & 2.02$^{+1.98}_{-0.73}$ & 2.88$\pm$0.89 & $-1.50$ & 1.35 & 0.39 & 0.06 \\
150418 & 232.66 & -3.23 & 776.20$\pm$0.50 & 1.76$^{+1.32}_{-0.99}$ & 0.80$\pm$0.30 & $-1.50$ & 1.35 & 0.39 & 0.06 \\
150610 & 278.00 & 16.50 & 1593.90$\pm$0.60 & 1.40$^{+0.75}_{-0.75}$ & 2.00$\pm$1.00 & $-1.50$ & 1.35 & 0.39 & 0.06 \\
150807 & 333.89 & -53.60 & 266.50$\pm$0.10 & 44.80$^{+8.40}_{-8.40}$ & 0.35$\pm$0.05 & $-7.00$ & 1.35 & 0.39 & 0.06 \\
151230 & 239.00 & 34.80 & 960.40$\pm$0.50 & 1.85$^{+0.26}_{-0.26}$ & 4.40$\pm$0.50 & $-1.50$ & 1.35 & 0.39 & 0.06 \\
160102 & 18.90 & -60.80 & 2596.10$\pm$0.30 & 1.70$^{+0.56}_{-0.56}$ & 3.40$\pm$0.80 & $-1.50$ & 1.35 & 0.39 & 0.06 \\
171209 & 332.20 & 6.24 & 1457.40$\pm$0.03 & 3.70$^{+0.10}_{-0.10}$ & 2.50$\pm$0.25 & $-1.50$ & 1.35 & 0.39 & 0.06 \\
180309 & 10.90 & -45.40 & 263.42$\pm$0.01 & 13.11$^{+0.26}_{-0.26}$ & 0.47$\pm$0.05 & $-1.50$ & 1.35 & 0.39 & 0.06 \\
180714 & 14.80 & 8.72 & 1467.92$\pm$0.30 & 1.74$^{+0.05}_{-0.05}$ & 2.90$\pm$0.29 & $-1.50$ & 1.35 & 0.39 & 0.06 \\ \hline
\multicolumn{10}{|c|}{Repeating FRB} \\ \hline
180908.J1232+74 & 124.70 & 42.90 & 195.70$\pm$0.90 & 2.70$^{+1.81}_{-1.81}$ & 1.91$\pm$0.10 & $-1.50$ & 0.60 & 0.02 & 0.98 \\
180916.J0158+65 & 129.70 & 3.70 & 349.20$\pm$0.40 & 2.50$^{+1.20}_{-1.20}$ & 1.30$\pm$0.07 & $-1.50$ & 0.60 & 0.39 & 1.00 \\
181017.J1705+68 & 99.20 & 34.80 & 1281.00$\pm$0.60 & 16.00$^{+8.11}_{-8.11}$ & 20.20$\pm$1.70 & $-1.50$ & 0.60 & 0.39 & 1.00 \\
181017.J18+81 & 113.30 & 27.80 & 301.55$\pm$0.25 & 2.75$^{+1.97}_{-1.97}$ & 3.10$\pm$0.95 & $-1.50$ & 0.60 & 0.02 & 0.98 \\
181119.J12+65 & 124.50 & 52.00 & 364.00$\pm$0.30 & 2.50$^{+1.20}_{-1.20}$ & 2.66$\pm$0.10 & $-1.50$ & 0.60 & 0.39 & 1.00 \\
181128.J0456+63 & 146.60 & 12.40 & 450.20$\pm$0.30 & 4.40$^{+2.20}_{-2.20}$ & 2.43$\pm$0.16 & $-1.50$ & 0.60 & 0.39 & 1.00 \\
190110.J1353+48 & 97.50 & 65.70 & 222.10$\pm$0.35 & 2.70$^{+1.34}_{-1.34}$ & 3.25$\pm$0.30 & $-1.50$ & 0.60 & 0.02 & 0.98 \\
190116.J1249+27 & 210.50 & 89.50 & 443.60$\pm$0.80 & 2.80$^{+1.40}_{-1.40}$ & 1.50$\pm$0.30 & $-1.50$ & 0.60 & 0.39 & 1.00 \\
190117.J2207+17 & 76.40 & -30.30 & 393.30$\pm$0.20 & 5.90$^{+2.33}_{-2.33}$ & 2.56$\pm$0.13 & $-1.50$ & 0.60 & 0.02 & 0.98 \\
190208.J1855+46 & 76.80 & 18.90 & 580.05$\pm$0.20 & 1.70$^{+0.88}_{-0.88}$ & 1.11$\pm$0.15 & $-1.50$ & 0.60 & 0.02 & 0.98 \\
190212.J02+20 & 148.10 & -38.70 & 651.50$\pm$0.40 & 3.00$^{+1.82}_{-1.82}$ & 4.00$\pm$0.40 & $-1.50$ & 0.60 & 0.02 & 0.98 \\
190222.J2052+69 & 104.90 & 15.90 & 460.20$\pm$0.25 & 5.45$^{+1.80}_{-1.80}$ & 2.71$\pm$0.85 & $-1.50$ & 0.60 & 0.39 & 1.00 \\
190417.J1939+59 & 91.50 & 17.40 & 1378.30$\pm$0.25 & 3.05$^{+1.31}_{-1.31}$ & 2.25$\pm$0.46 & $-1.50$ & 0.60 & 0.02 & 0.98 \\
190604.J1435+53 & 93.80 & 57.60 & 552.65$\pm$0.20 & 5.00$^{+2.48}_{-2.48}$ & 2.10$\pm$0.45 & $-1.50$ & 0.60 & 0.02 & 0.98 \\ \hline
    \end{tabular}\\
    Column (1) FRB ID. 
    (2) Galactic longitude.
    (3) Galactic latitude.
    (4) Observed dispersion measure. 
    (5) Observed fluence. 
    If the uncertainty, $\delta E_{\nu_{\rm obs}}$, is not provided, we calculated it as $\delta E_{\nu_{\rm obs}}$=$E_{\nu_{\rm obs}} \times \sqrt{(\delta w_{\rm obs}/w_{\rm obs})^{2}+(\delta S_{\rm obs}/S_{\rm obs})^{2}}$, where $S_{\rm obs}$ and $\delta S_{\rm obs}$ are the observed flux density and its uncertainty, respectively. 
    (6) Observed duration. 
    (7) Spectral index, i.e., $E_{\nu_{\rm obs}} \propto \nu_{\rm obs}^{\alpha}$, compiled from the FRBCAT \citep{Petroff2016}. 
    We assumed $\alpha=-1.5$ derived from a stacked spectrum of ASKAP FRBs \citep{Macquart2019}, if $\alpha$ is not available. 
    (8) Observed frequency. 
    (9) Channel width. 
    (10) Observational sampling time interval. 
    Each source of repeating FRBs has multiple measurements of each observed parameter due to its repetition.
    Median values over the multiple measurements are presented using individual repeating bursts which satisfy the selection criteria described in Section \ref{selection}.\\
    \end{flushleft}
\end{table*}

\begin{table*}
	\centering
	\caption{
	Derived parameters of FRBs used for the calculations of their luminosity functions.}
	\label{tabA2}
	\begin{flushleft}
	\begin{tabular}{|l|c|c|c|c|c|c|}\hline
\multicolumn{7}{|c|}{Non-repeating FRB} \\
(1)& (2)            & (3)       & (4)             & (5)                   & (6)                   & (7)               \\ 
ID & DM$_{\rm MW}$  & Redshift  & log$L_{\nu}$    & $\log \rho_{\rm original}$           & $\log \rho_{\rm keep N}$ & $\log \rho_{\rm optimised}$     \\
   & (pc cm$^{-3}$) &           & (erg Hz$^{-1}$) & (Gpc$^{-3}$ yr$^{-1}$)    & (Gpc$^{-3}$ yr$^{-1}$) & (Gpc$^{-3}$ yr$^{-1}$) \\ \hline 
010125 & 74.08 & 0.69$\pm$0.20 & 31.50$\pm$0.44 & 3.22$^{+0.45}_{-0.06}$ & 3.20$^{+0.50}_{-0.05}$ & 3.34$^{+0.38}_{-0.04}$ \\
010312 & 54.72 & 1.13$\pm$0.24 & 32.31$\pm$0.19 & 2.94$^{+0.14}_{-0.21}$ & 2.91$^{+0.19}_{-0.20}$ & 2.94$^{+0.16}_{-0.21}$ \\
010621 & 229.30 & 0.47$\pm$0.17 & 31.14$\pm$0.52 & 3.56$^{+0.47}_{-0.10}$ & 3.51$^{+0.53}_{-0.08}$ & 3.57$^{+0.33}_{-0.44}$ \\
010724 & 32.63 & 0.28$\pm$0.13 & 32.35$\pm$0.38 & 3.92$^{+0.28}_{-0.01}$ & 4.10$^{+0.28}_{-0.003}$ & 3.51$^{+0.04}_{-0.07}$ \\
090625 & 25.47 & 0.86$\pm$0.22 & 31.59$\pm$0.40 & 2.98$^{+0.20}_{-0.33}$ & 3.02$^{+0.24}_{-0.41}$ & 3.03$^{+0.20}_{-0.39}$ \\
110214 & 21.06 & 0.04$\pm$0.07 & 30.18$\pm$1.19 & 4.08$^{+0.02}_{-0.02}$ & 4.08$^{+0.03}_{-0.02}$ & 4.08$^{+0.02}_{-0.01}$ \\
110220 & 24.12 & 0.91$\pm$0.22 & 32.16$\pm$0.22 & 2.99$^{+0.06}_{-0.38}$ & 3.03$^{+0.22}_{-0.04}$ & 3.05$^{+0.06}_{-0.42}$ \\
110703 & 23.08 & 1.08$\pm$0.23 & 31.80$\pm$0.41 & 3.17$^{+0.22}_{-0.23}$ & 3.11$^{+0.42}_{-0.30}$ & 3.17$^{+0.23}_{-0.27}$ \\
121002 & 60.02 & 1.60$\pm$0.27 & 32.21$\pm$0.37 & 2.71$^{+0.32}_{-0.16}$ & 2.69$^{+0.33}_{-0.15}$ & 2.71$^{+0.32}_{-0.16}$ \\
130626 & 64.76 & 0.87$\pm$0.21 & 31.43$\pm$0.40 & 2.98$^{+0.38}_{-0.25}$ & 3.02$^{+0.37}_{-0.37}$ & 3.04$^{+0.41}_{-0.33}$ \\
130628 & 46.93 & 0.37$\pm$0.14 & 30.52$\pm$0.38 & 3.68$^{+0.35}_{-0.19}$ & 3.60$^{+0.54}_{-0.20}$ & 3.54$^{+0.33}_{-0.40}$ \\
130729 & 25.42 & 0.82$\pm$0.21 & 31.74$\pm$0.48 & 3.09$^{+0.43}_{-0.07}$ & 3.08$^{+0.54}_{-0.17}$ & 3.17$^{+0.44}_{-0.06}$ \\
131104 & 218.01 & 0.52$\pm$0.17 & 31.14$\pm$0.38 & 3.21$^{+0.45}_{-0.08}$ & 3.18$^{+0.40}_{-0.06}$ & 3.34$^{+0.30}_{-0.13}$ \\
140514 & 24.16 & 0.50$\pm$0.17 & 30.84$\pm$0.44 & 3.72$^{+0.36}_{-0.11}$ & 3.65$^{+0.45}_{-0.11}$ & 3.58$^{+0.40}_{-0.45}$ \\
150215 & 194.79 & 0.90$\pm$0.22 & 31.60$\pm$0.34 & 2.99$^{+0.25}_{-0.28}$ & 3.03$^{+0.26}_{-0.34}$ & 3.04$^{+0.27}_{-0.33}$ \\
150418 & 306.33 & 0.42$\pm$0.16 & 30.81$\pm$0.44 & 3.33$^{+0.63}_{-0.13}$ & 3.30$^{+0.69}_{-0.11}$ & 3.55$^{+0.14}_{-0.49}$ \\
150610 & 120.93 & 1.50$\pm$0.27 & 31.93$\pm$0.29 & 2.77$^{+0.34}_{-0.20}$ & 2.75$^{+0.44}_{-0.22}$ & 2.77$^{+0.34}_{-0.20}$ \\
150807 & 25.48 & 0.16$\pm$0.10 & 30.93$\pm$0.64 & 3.88$^{+0.03}_{-0.05}$ & 4.06$^{+0.03}_{-1.24}$ & 3.46$^{+0.04}_{-0.04}$ \\
151230 & 37.76 & 0.91$\pm$0.22 & 31.57$\pm$0.23 & 2.99$^{+0.36}_{-0.19}$ & 3.03$^{+0.34}_{-0.15}$ & 3.05$^{+0.38}_{-0.17}$ \\
160102 & 21.78 & 2.74$\pm$0.37 & 32.56$\pm$0.19 & 2.47$^{+0.11}_{-0.13}$ & 2.51$^{+0.08}_{-0.15}$ & 2.47$^{+0.11}_{-0.13}$ \\
171209 & 183.72 & 1.29$\pm$0.25 & 32.20$\pm$0.18 & 2.48$^{+0.14}_{-0.08}$ & 2.47$^{+0.13}_{-0.07}$ & 2.48$^{+0.14}_{-0.08}$ \\
180309 & 29.99 & 0.15$\pm$0.10 & 30.71$\pm$0.45 & 3.87$^{+0.03}_{-0.05}$ & 4.06$^{+0.03}_{-1.34}$ & 3.46$^{+0.05}_{-0.04}$ \\
180714 & 188.83 & 1.29$\pm$0.25 & 31.88$\pm$0.18 & 2.91$^{+0.18}_{-0.14}$ & 2.88$^{+0.19}_{-0.13}$ & 2.91$^{+0.19}_{-0.14}$ \\ \hline
\multicolumn{7}{|c|}{Repeating FRB} \\ \hline
180908.J1232+74 & 30.62 & 0.07$\pm$0.07 & 28.74$\pm$0.60 & 3.00$^{+0.27}_{-0.03}$ & 3.01$^{+0.28}_{-0.03}$ & 3.00$^{+0.29}_{-0.03}$ \\
180916.J0158+65 & 309.41 & 0.034 & 28.11$\pm$0.21 & 3.79$^{+0.37}_{-0.24}$ & 3.83$^{+0.39}_{-0.23}$ & 3.79$^{+0.37}_{-0.23}$ \\
181017.J1705+68 & 36.84 & 1.25$\pm$0.26 & 32.29$\pm$0.30 & 0.05$^{+0.08}_{-0.36}$ & 0.10$^{+0.05}_{-0.42}$ & 0.01$^{+0.08}_{-0.31}$ \\
181017.J18+81 & 48.31 & 0.17$\pm$0.10 & 29.63$\pm$0.54 & 1.93$^{+0.50}_{-0.11}$ & 1.93$^{+0.50}_{-0.12}$ & 1.93$^{+0.51}_{-0.26}$ \\
181119.J12+65 & 25.71 & 0.28$\pm$0.13 & 30.02$\pm$0.46 & 1.44$^{+0.58}_{-0.44}$ & 1.44$^{+0.56}_{-0.47}$ & 1.63$^{+0.48}_{-0.16}$ \\
181128.J0456+63 & 148.17 & 0.23$\pm$0.13 & 30.11$\pm$0.50 & 1.38$^{+0.41}_{-0.70}$ & 1.40$^{+0.41}_{-0.68}$ & 1.41$^{+0.52}_{-0.12}$ \\
190110.J1353+48 & 21.83 & 0.11$\pm$0.09 & 29.20$\pm$0.55 & 2.51$^{+0.42}_{-0.08}$ & 2.51$^{+0.41}_{-0.08}$ & 2.51$^{+0.45}_{-0.11}$ \\
190116.J1249+27 & 19.54 & 0.37$\pm$0.15 & 30.36$\pm$0.43 & 1.05$^{+0.45}_{-0.13}$ & 1.05$^{+0.44}_{-0.11}$ & 0.94$^{+0.50}_{-0.07}$ \\
190117.J2207+17 & 39.86 & 0.29$\pm$0.14 & 30.45$\pm$0.43 & 1.40$^{+0.26}_{-0.02}$ & 1.42$^{+0.26}_{-0.01}$ & 0.79$^{+0.44}_{-0.04}$ \\
190208.J1855+46 & 61.45 & 0.48$\pm$0.16 & 30.37$\pm$0.40 & 0.95$^{+0.51}_{-0.18}$ & 0.95$^{+0.52}_{-0.15}$ & 0.87$^{+0.54}_{-0.24}$ \\
190212.J02+20 & 35.46 & 0.58$\pm$0.19 & 30.82$\pm$0.42 & 0.75$^{+0.55}_{-0.17}$ & 0.74$^{+0.57}_{-0.26}$ & 1.64$^{+0.11}_{-0.03}$ \\
190222.J2052+69 & 92.82 & 0.31$\pm$0.14 & 30.43$\pm$0.40 & 1.16$^{+0.43}_{-0.10}$ & 1.15$^{+0.45}_{-0.09}$ & 1.00$^{+0.47}_{-0.08}$ \\
190417.J1939+59 & 73.58 & 1.32$\pm$0.26 & 31.57$\pm$0.26 & 0.10$^{+0.22}_{-0.25}$ & 0.11$^{+0.16}_{-0.28}$ & 0.06$^{+0.20}_{-0.21}$ \\
190604.J1435+53 & 23.83 & 0.49$\pm$0.17 & 30.75$\pm$0.42 & 0.74$^{+0.35}_{-0.25}$ & 0.78$^{+0.31}_{-0.34}$ & 0.79$^{+0.29}_{-0.42}$ \\ \hline
    \end{tabular}\\
    Column (1) FRB ID. (2) Dispersion measure of inter stellar medium in the Milky Way along a line of sight to FRB, based on YMW16 model \citep{Yao2017}. 
    DM$_{\rm MW}$ is accumulated up to 10 kpc. 
    (3) Redshift calculated from dispersion measure (see Section \ref{calcz}) or spectroscopic redshift for FRB180916.J0158+65 \citep{Marcote2020}. 
    (4) Time-integrated luminosity at the rest frame 1.83 GHz. 
    (5) Individual volumetric occurrence rate ($\rho$ in Eq. \ref{eqeqrhosum}) calculated from the redshift bins of $0.01<z\leq0.35$, $0.35<z\leq1.0$, and $1.0<z\leq3.0$ for non-repeating FRBs and $0.01<z\leq0.3$, $0.3<z\leq0.7$, and $0.7<z\leq1.5$ for repeating FRBs.
    (6) Same as (5) except that the values are calculated from the redshift bins of $0.02<z\leq0.3$, $0.3<z\leq0.95$, and $0.95<z\leq2.8$ for non-repeating FRBs and $0.02<z\leq0.295$, $0.295<z\leq0.65$, and $0.65<z\leq1.4$ for repeating FRBs.
    (7) Same as (5) except that the values are calculated from the redshift bins of $0.01<z\leq0.5$, $0.5<z\leq1.0$, and $1.0<z\leq3.0$ for non-repeating FRBs and $0.01<z\leq0.2$, $0.2<z\leq0.58$, and $0.58<z\leq1.5$ for repeating FRBs.
    Uncertainties of $\rho$ include observational uncertainties of fluence, dispersion measure, and pulse duration together with the fluctuation of DM$_{\rm IGM}$ estimated from a simulation \citep{Zhu2018} (see Section \ref{calcz} and APPENDIX B for details).
    The redshift uncertainty of FRB180916.J0158+65 is approximated as 0.0, since its spectroscopic redshift is available \citep{Marcote2020}.
    Each source of repeating FRBs has multiple measurements of each observed parameter due to its repetition. 
    The redshift and $L_{\nu}$ are individually calculated for repeating bursts which satisfy the selection criteria described in Section \ref{selection}.
    Median values over the individually calculated redshift and $L_{\nu}$ are presented for each repeating FRB. 
    These median values of repeating FRBs are used for the calculation of their $\rho$.
    \end{flushleft}
\end{table*}

\section{Line-of-sight fluctuation of intergalactic medium}
\begin{figure}
    \centering
    \includegraphics[width=1.0\columnwidth]{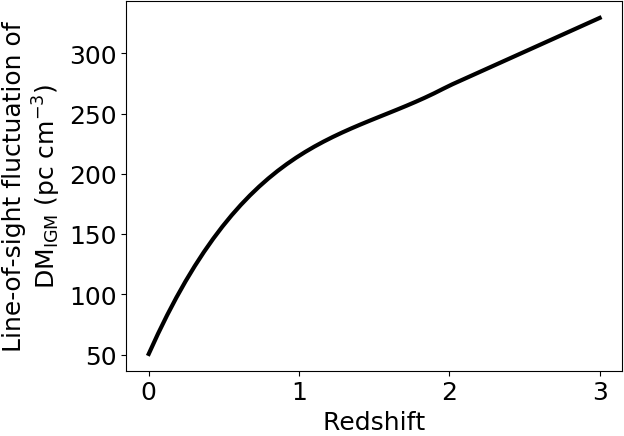}
    \caption{
    The fluctuation of dispersion measure of intergalactic medium along different lines of sight assumed in this work.
    The polynomial function of the third degree is fitted to the result of cosmological simulation \citep{Zhu2018} at $0<z\leq2$. 
    The linear function is fitted between $1<z\leq2$ to extrapolate the fluctuation up to $z=3$. 
    The best-fit functions are described in Eqs. \ref{sigmaDM1} and \ref{sigmaDM2}.
    }
    \label{figB1}
\end{figure}

We describe the fluctuation of the dispersion measure of the intergalactic medium along different lines of sight.
In this work, we conservatively use the largest fluctuation of dispersion measure, $\sigma_{\rm DM_{\rm IGM}}$, among the cosmological simulations by \citet{Zhu2018}.
For computational purposes, a polynomial function of the third degree was fitted to $\sigma_{\rm DM_{\rm IGM}}$ as a function of redshift between $0<z\leq2$, where $\sigma_{\rm DM_{\rm IGM}}$ is simulated in \citet{Zhu2018}.
We extrapolated $\sigma_{\rm DM_{\rm IGM}}$ up to $z=3$ by fitting a linear function to $\sigma_{\rm DM_{\rm IGM}}$ at $1<z\leq2$.
The best-fit functions are 
\begin{empheq}[left={\sigma_{\rm DM_{\rm IGM}}=\empheqlbrace}]{alignat=2}
\label{sigmaDM1}
&50.52 +280.27z -147.28z^{2} +31.42z^{3}&(0<z\leq2)\\
\label{sigmaDM2}
&160.76 +56.20z&(2<z\leq3).
\end{empheq}
These functions are presented in Fig. \ref{figB1}.

\section{Redshift cut and bins}
\begin{figure*}
    \centering
    \includegraphics[width=1.0\columnwidth]{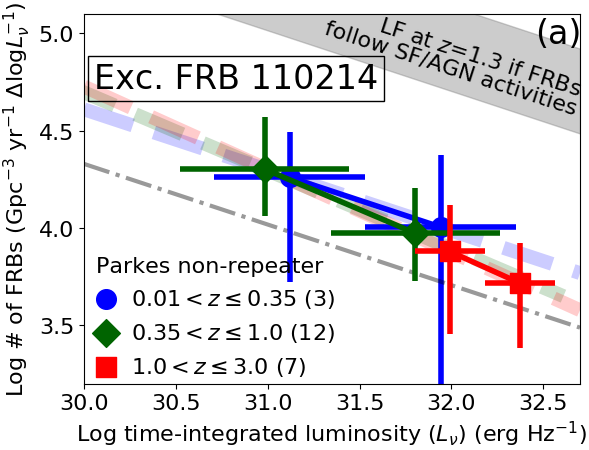}
    \includegraphics[width=1.0\columnwidth]{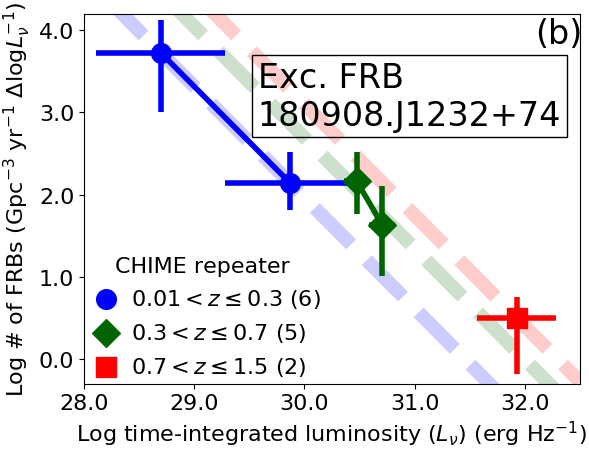}
    \caption{
    Same as Fig. \ref{fig4} except for excluding low-$z$ FRBs, non-repeating FRB 110214 at $z=0.04\pm0.07$ and repeating FRB 180908.J1232+74 at $z=0.07\pm0.07$.
    }
    \label{figC1}
\end{figure*}

As mentioned in Section \ref{zcut}, the redshift cut is applied to the FRB samples before simulating the redshift uncertainty.
Therefore, low-$z$ FRBs, including non-repeating FRB 110214 at $z=0.04\pm0.07$ and repeating FRB 180908.J1232+74 at $z=0.07\pm0.07$, sometimes go out of the redshift cut within the uncertainties. 
Here we examine how this uncertainty affects the luminosity functions.
We performed the same analyses described in Section \ref{calcLF} by excluding these FRBs from our sample.
The derived luminosity functions are shown in Fig. \ref{figC1}.
We found no significant difference between luminosity functions including/excluding these low-$z$ FRBs in the sense that non-repeating FRBs do not show significant redshift evolution of their luminosity functions, while those of repeating FRBs increase towards higher redshifts.

The derived luminosity functions in Section \ref{calcLF} could be sensitive to how redshift bins are decided, since $1/V_{\rm max,4\pi}$ of each FRB becomes extremely large if $d_{\rm max}$ of the FRB is very close to the lower bound, $d_{\rm min}$ (see Eq. \ref{vmax}). 
Such FRBs may affect the shapes of luminosity functions.
To check this effect, we test two additional cases of redshift bins: (i) changing the redshift bins while keeping the sample size the same in each bin, and (ii) optimising the redshift bins so that the averaged value of $V_{4\pi}/V_{\rm max,4\pi}$ in each bin approaches 0.5 \citep[e.g., ][]{Avni1980,Jurek2013}, where $V_{4\pi} = 4\pi/3(d_{\rm FRB}^{3}-d_{\rm min}^{3})$ and $d_{\rm FRB}$ is the comoving distance to each FRB.

\begin{figure*}
    \centering
    \includegraphics[width=2.0\columnwidth]{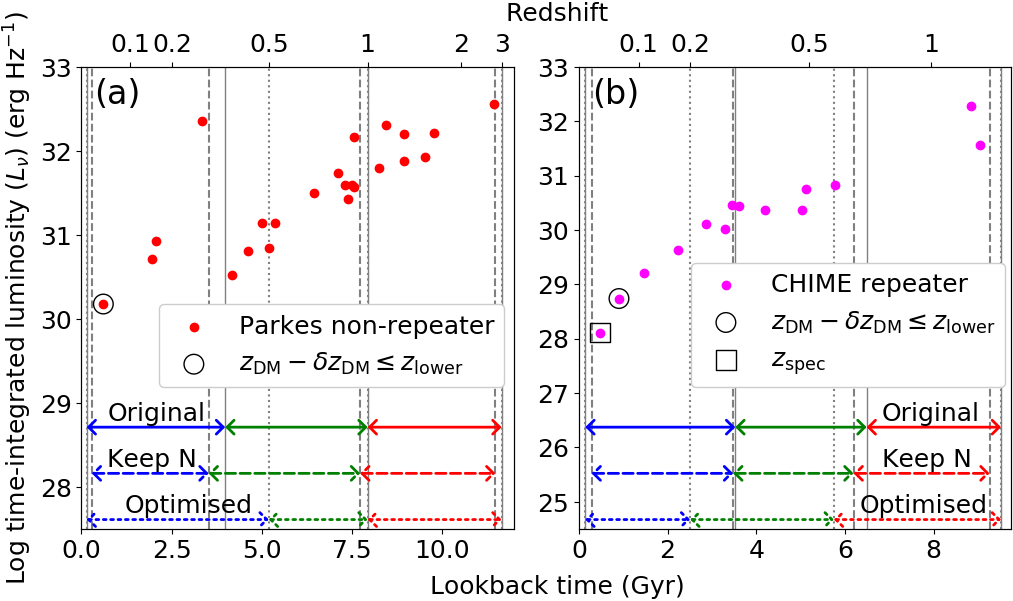}
    \caption{
    Time-integrated luminosity as a function of lookback time (or redshift) of FRBs selected in Section \ref{selection}.
    (a) Parkes non-repeating FRBs are shown.
    The vertical solid lines are originally adopted redshift bins in Section \ref{calcLF} such that their time intervals are equal, corresponding to $\sim$ 4 Gyr.
    The vertical dashed and dotted lines are additionally considered redshift bins, i.e., (i) changing the redshift bins while keeping the sample size the same in each bin, and (ii) optimising the redshift bins so that the averaged value of $V_{4\pi}/V_{\rm max_4\pi}$ in each bin approaches 0.5, respectively.
    FRB 110214 at $z=0.04\pm0.07$ is marked by a circle, because this source could be out of the redshift cut ($z\geq0.01$) within its uncertainty.
    (b) Same as (a) except for CHIME repeating FRBs.
    FRB 180908.J1232+74 at $z=0.07\pm0.07$ is marked by a circle, because this source could be out of the redshift cut ($z\geq0.01$) within its uncertainty.
    FRB 180916.J0158+65 at $z=0.034$ is marked by a square, because the spectroscopic redshift is measured for this source \citep{Marcote2020}.
    }
    \label{figC2}
\end{figure*}

In case (i), we adopt $0.02<z\leq0.3$, $0.3<z\leq0.95$, and $0.95<z\leq2.8$ for non-repeating FRBs, and $0.02<z\leq0.295$, $0.295<z\leq0.65$, and $0.65<z\leq1.4$ for repeating FRBs.
The number of sample in each redshift bin is the same as the original one so that we can investigate the bin-edge effect.
In case (ii), we changed the redshift bins more flexibly allowing the number of sample in each bin to change, and calculated $V_{4\pi}/V_{\rm max,4\pi}$.
The $V_{\rm max}$ method assumes no redshift evolution within each redshift bin, where the sample mean of $V_{4\pi}/V_{\rm max,4\pi}$ is 0.5 if the assumption holds \citep[e.g., ][]{Avni1980,Jurek2013}.
We searched for the optimal redshift bins which shows the mean of $V_{4\pi}/V_{\rm max,4\pi}$ close to 0.5 at each redshift bin.
The optimised redshift bins of non-repeating FRBs are $0.01<z\leq0.5$, $0.5<z\leq1.0$, and $1.0<z\leq3.0$, where the mean values of $V_{4\pi}/V_{\rm max,4\pi}$ are 0.4, 0.6, and 0.5 respectively.
The optimised redshift bins of repeating FRBs are $0.01<z\leq0.2$, $0.2<z\leq0.58$, and $0.58<z\leq1.5$, where the mean values of $V_{4\pi}/V_{\rm max,4\pi}$ are 0.7, 0.5, and 0.5 respectively.
These mean values are closer to 0.5 than the values in original redshift bins, i.e., 0.2 (0.7), 0.6 (0.5), and 0.5 (0.7) in the low, middle, and high-$z$ redshift bins for non-repeating (repeating) FRBs, respectively.
These two cases of redshift bins are summarised in Fig. \ref{figC2} along with the original redshift bins in Section \ref{calcLF}.

\begin{figure*}
    \centering
    \includegraphics[width=1.0\columnwidth]{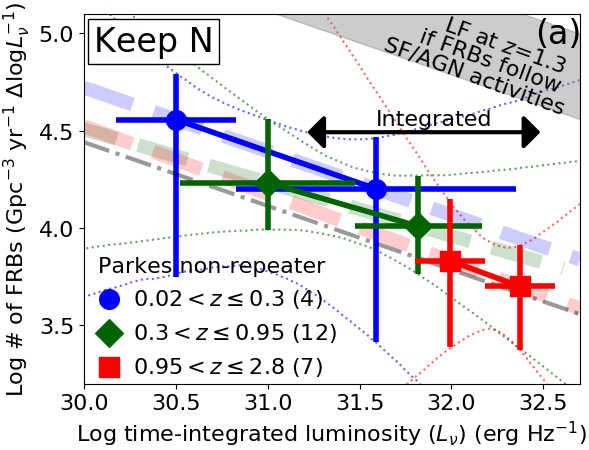}
    \includegraphics[width=1.0\columnwidth]{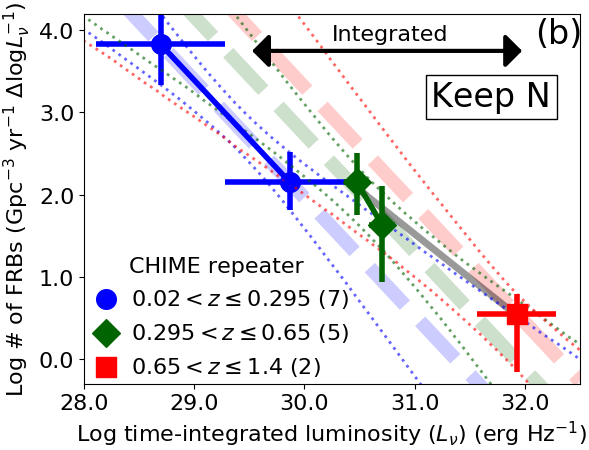}
    \includegraphics[width=1.0\columnwidth]{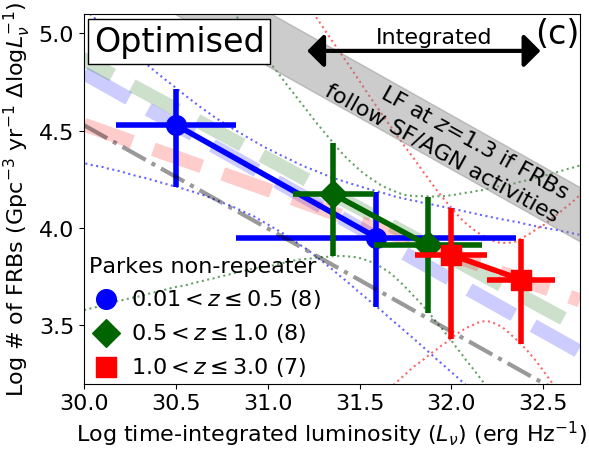}
    \includegraphics[width=1.0\columnwidth]{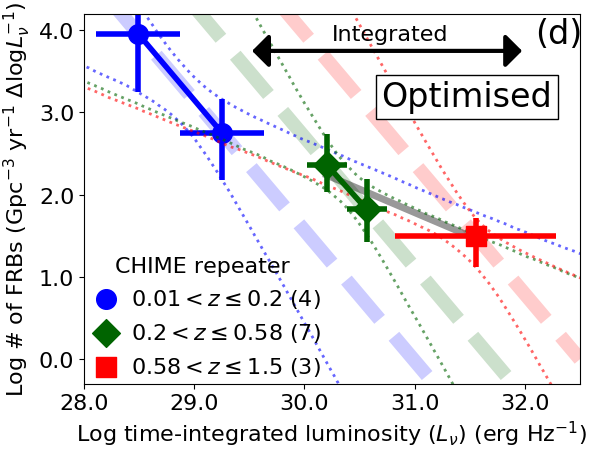}
    \caption{
    Same as Fig. \ref{fig4} except for different redshift bins: (a) and (b) changing the redshift bins while keeping the sample size the same in each bin, and (c) and (d) optimising the redshift bins so that the averaged value of $V_{4\pi}/V_{\rm max_4\pi}$ in each bin approaches 0.5.
    In panel (c), the grey shaded region predicts a moderate redshift evolution compared to Figs. \ref{fig4}a, \ref{figC1}a and \ref{figC3}a, since the median redshift in the low-$z$ bin ($0.01<z\leq0.5$, $z_{\rm median}=0.3$) is higher than that of other redshift-bin cases ($z_{\rm median}=0.16$).
    }
    \label{figC3}
\end{figure*}

\begin{figure*}
    \centering
    \includegraphics[width=2.0\columnwidth]{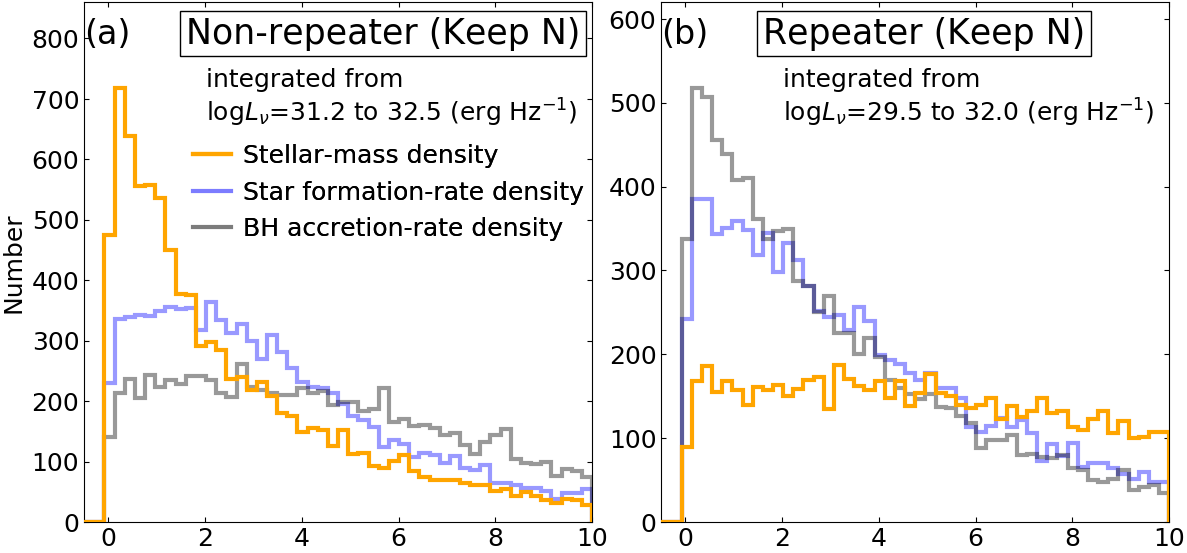}
    \includegraphics[width=2.0\columnwidth]{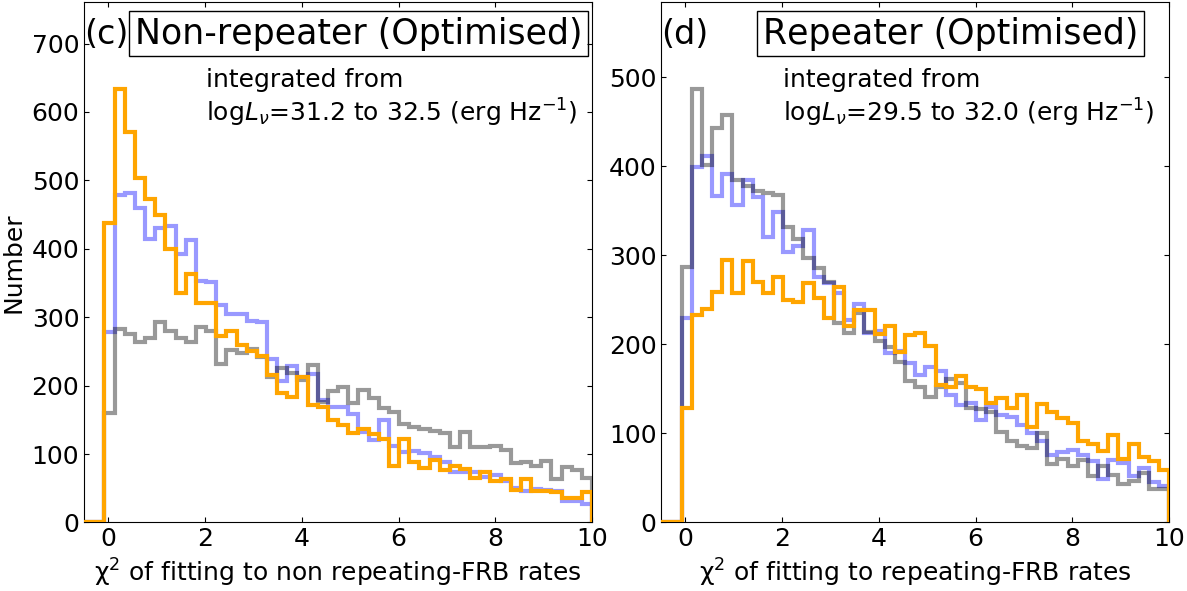}
    \caption{
    Same as panels (a) and (b) in Fig. \ref{fig6} except for different redshift bins: (a) and (b) changing the redshift bins while keeping the sample size the same in each bin, and (c) and (d) optimising the redshift bins so that the averaged value of $V_{4\pi}/V_{\rm max_4\pi}$ in each bin approaches 0.5.
    }
    \label{figC4}
\end{figure*}

Based on these redshift bins, we performed the same analyses in Section \ref{analysis} and \ref{results}. 
The derived luminosity functions and $\chi^{2}$ distributions are shown in Figs. \ref{figC3} and \ref{figC4}, respectively.
We found no significant difference in the luminosity functions in the sense that non-repeating FRBs do not show significant redshift evolution of their luminosity functions while those of repeating FRBs increase towards higher redshifts if their slope is constant over $z=0.01$ to 1.5. 
In both redshift-bin cases, the cosmic stellar-mass density shows better fits to the volumetric occurrence rates of non-repeating FRBs than the cosmic star formation-rate density.
The cosmic star formation-rate density or BH accretion-rate density shows better fits to the volumetric occurrence rates of repeating FRBs.
We conclude that the different redshift bins do not affect our results in Section \ref{results} significantly, as far as equal time intervals or optimised redshift bins are utilised.

\bsp	
\label{lastpage}
\end{document}